\documentclass[a4paper,fleqn,usenatbib]{mnras}

\usepackage{newtxtext,newtxmath}

\usepackage[T1]{fontenc}
\usepackage{ae,aecompl}

\usepackage{graphicx}	
\usepackage{amsmath}	
\usepackage{amssymb}	
\usepackage{adjustbox}
\usepackage{tablefootnote}
\usepackage{natbib}
\usepackage{epsf}
\usepackage{color}
\usepackage{amsmath}
\usepackage{subfigure}
\usepackage{footnote}
\usepackage{graphicx}

\title [SAMI Kinematics
  of Dusty Early-Type Galaxies]{The SAMI Galaxy Survey: Kinematics of Dusty Early-Type Galaxies}

\author[R. Bassett et al.]{
R. Bassett$^1$\thanks{E-mail: robert.bassett@uwa.edu.au (ICRAR)},
K. Bekki$^1$,
L. Cortese$^1$, 
W. J. Couch$^{2}$, 
A. E. Sansom$^{3}$, 
J. van de Sande$^{4}$,\newauthor
J. J. Bryant$^{2,4,5}$,
C. Foster$^{2}$,
S. M. Croom$^{4,5}$,
S. Brough$^{2}$,
S. M. Sweet$^{6}$,
A. M. Medling$^{6,7,8}$,\newauthor
M. S. Owers$^{2,9}$,
S. P. Driver$^{1}$,
L. J. M. Davies$^{1}$,
O. I. Wong$^{1,5}$,
B. A. Groves$^{6}$,\newauthor
J. Bland-Hawthorn$^{4}$,
S. N. Richards$^{2,4,5}$,
M. Goodwin$^{2}$,
I. S. Konstantopoulos$^{2,10}$,\newauthor
J. S. Lawrence$^{2}$
\\
$^{1}$International Centre for Radio Astronomy Research,
  University of Western Australia, 7 Fairway, Crawley, WA 6009,
  Australia\\
$^{2}$Australian Astronomical Observatory, PO Box 915, North Ryde,
  NSW 1670, Australia\\
$^{3}$Jeremiah Horrocks Institute, University of Central
  Lancashire, Preston PR1 2HE, UK\\
$^{4}$Sydney Institute for Astronomy, School of Physics,
  A28, The University of Sydney, NSW 2006, Australia\\
$^{5}$ARC Centre of Excellence for All-sky Astrophysics
  (CAASTRO)\\
$^{6}$Research School for Astronomy and Astrophysics, Australian
  National University, Canberra, ACT 2611, Australia\\
$^{7}$Cahill Center for Astronomy and Astrophysics, California
Institute of Technology, MS 249-17 Pasadena, CA 91125, USA\\
$^{8}$Hubble Fellow\\
$^{9}$Department of Physics and Astronomy, Macquarie University,
  NSW 2109, Australia\\
$^{10}$Envizi Suite 213, National Innovation Centre, Australian
  Technology Park, 4 Cornwallis Street, Eveleigh NSW 2015, Australia
}

\date{Accepted XXX. Received YYY; in original form ZZZ}

\pubyear{2016}

\begin{document}
\label{firstpage}
\pagerange{\pageref{firstpage}--\pageref{lastpage}}
\maketitle

\begin{abstract}

Recently, large samples of visually classified early-type galaxies
(ETGs) containing dust
have been identified using
space-based infrared observations with the
\textit{Herschel} Space Telescope. The presence of large quantities of
dust in massive ETGs is peculiar as
X-ray halos of these galaxies are expected to destroy dust in $\sim$10$^{7}$ yr (or less). This
has sparked a debate regarding the origin of the dust: is it
internally produced by asymptotic giant branch (AGB) stars, or
is it accreted externally through mergers? We
examine the 2D stellar and ionised gas kinematics of dusty ETGs using IFS observations from the SAMI galaxy
survey, and integrated star-formation rates, stellar masses,
and dust masses from the GAMA survey. Only 8\% (4/49) of
visually-classified ETGs are
kinematically consistent with being dispersion-supported
systems. These
``dispersion-dominated galaxies'' exhibit discrepancies between
stellar and ionised gas kinematics, either offsets in the
kinematic position angle or large differences in the
rotational velocity, and are 
outliers in star-formation rate at a fixed dust mass
compared to normal star-forming galaxies. These
properties are suggestive of recent
merger activity. The remaining $\sim$90\% of dusty ETGs have low
velocity dispersions and/or large circular velocities, typical of
``rotation-dominated galaxies''. These
results, along with the general evidence of published works on X-ray
emission in ETGs, suggest that they are unlikely to host hot, X-ray
gas consistent with their low $M_{*}$ when compared to
dispersion-dominated galaxies. This means dust will be long lived and thus these galaxies do not
require external scenarios for the origin of their dust content. 

\end{abstract}




\section{Introduction}\label{section:intro}

The recent launch of the \textit{Herschel} Space Telescope has made
it possible for astronomers to study cold dust in a wide variety of
galaxies with unprecedented sensitivity. As a consequence, a number of
teams have identified large samples of visually-classified early-type
galaxies (ETGs) that clearly harbour massive reservoirs of cold dust
\citep{cortese12b,rowlands12,smith12,diserego13,agius13,agius15,dariush16}. Although
dust is closely related to the formation of stars in star-forming,
late-type galaxies (LTGs), this may not be the case in ETGs where the
level of on-going star formation is typically much lower (if not non-existent). Furthermore,
massive ETGs are known to contain large amounts of hot, X-ray emitting
gas that is inhospitable to fragile dust grains.
This hot gas rapidly destroys dust through a process known as thermal
sputtering, resulting in a dust lifetime of $\sim$10$^{5}$-10$^{7}$ yr
\citep{draine79,itoh89,tsai95,mathews03,clemens10,anderson15}. 

The now undisputed presence of large quantities of dust in some ETGs has
sparked a debate as to its origins. Many works have suggested that
dust found in ETGs must have been recently accreted via mergers
with gas-rich satellites
\citep{goud98,gomez10,kaviraj09,davis11,shabala12,kaviraj13,davis14,dariush16}. In such a merger, the accreted dust
will be embedded in a cold medium (either atomic or molecular gas)
that can provide shielding from X-ray photons, resulting in a longer lifetime than for
dust produced internally
\citep{temi07a,clemens10,dasyra12,finkelman12}. Alternatively,
the dust may result from internal processes such as cooling of hot
halo gas \citep{fabian94,bregman05,lagos14} or production in
asymptotic giant-branch (AGB) stars \citep{knapp85,knapp92,athey02,matsuura09,nanni13}. Currently, there is no clear
consensus regarding the internal versus external origins of dust in
ETGs, and it is possible that both play some role with the balance
between the two sources depending on the properties of individual
galaxies \citep{rampazzo05,cappe11,finkelman12}.

Much of the recent work on dusty ETGs is based on samples selected
by visual morphology. In such cases it is not clear how certain we can
be that such galaxies
host a hot, X-ray emitting halo. The X-ray properties of ETGs vary
considerably. This X-ray emission is less dominant in lower
mass ETGs \citep[e.g.][]{boroson11}, in (apparently) younger ETGs
\citep{sansom00,sansom06}, and in ETGs with higher star formation
\citep{su15}. Environment is also thought to play a role
\citep[e.g.][]{mulchaey10}. Clear evidence of diffuse X-ray emission is found in 
massive galaxy clusters as well as the most massive individual
ETGs \citep[10$^{10.8}$ M$_{\odot}$ and higher; e.g.][]{anderson15}
with X-ray luminosities ($L_{X}$) significantly larger than 10$^{40}$
ergs s$^{-1}$. For LTGs, \citet{mineo12} finds $L_{X}$< 10$^{40}$ ergs
s$^{-1}$ corresponding roughly to the high $L_{X}$ cut-off for X-ray binary stars \citep[see][for a
review]{fabbiano06}. Thus, for galaxies observed with $L_{X}$ <
$\sim$10$^{40}$ ergs s$^{-1}$, particularly those with recent star
formation, X-ray emission can be attributed to the
cumulative emission from supernova remnants and X-ray binaries. 

Recent spectroscopic work has provided a connection between
galaxy kinematics and X-ray properties.
In particular, galaxies with stellar
velocity dispersions ($\sigma$) larger than $\sim$150 km s$^{-1}$ are
often found to have X-ray luminosities in excess of 10$^{40}$ ergs
s$^{-1}$ \citep{boroson11,sarzi13,kim15,goulding16}, and these galaxies appear to
extend the relationship between $L_{X}$ and stellar mass
found in massive galaxy clusters \citep[e.g.][and references
therein]{wu99,oritzgil04,zhang11} to lower mass systems. Below
$\sigma$ = 150 km s$^{-1}$, all ETGs studied by \citet{goulding16}
have $L_{X}$ < 10$^{40}$ ergs s$^{-1}$; in the range attributed to
X-ray binaries by \citet{mineo12}. Furthermore, recent simulations by
\citet{negri14b} have shown that galaxy rotation can also act to
reduce $L_{X}$. This occurs because conservation of angular momentum
in rotating galaxy models encourages the growth of cold gas disks, preventing
large amounts of hot gas from collecting in the central region. These results suggest that kinematic
observations of visually-selected, dusty ETGs may distinguish
galaxies embedded in a massive halo of hot gas from those more
hospitable to long lived dust reservoirs. It is also worth noting that visual morphology and kinematic classifications
are not always well correlated \citep[see][for a recent review of this topic]{cappe16}.

This connection between X-ray emitting gas content and kinematics
shows that 
spatially-resolved observations using integral field spectroscopy
(IFS), which give a detailed description of a galaxy's kinematics, can help in understanding the
origins of dust in ETGs. IFS observations of large samples of galaxies
identified as dusty ETGs provide a step forward in two
respects. First,
because IFS allows global measurements of stellar $\sigma$ covering most of
the galaxy,
they can clearly identify galaxies with large stellar $\sigma$ that most
likely host X-ray emitting gas. Second, IFS
observations provide a strong indicator of recent merger activity
through the direct comparison of ionised gas and stellar
kinematics. The work of \citet{davis11} using galaxies from the
ATLAS$^{\textrm{3D}}$ survey is an example in this vein,
showing a connection between the detection of molecular gas in ETGs
and misalignments between ionised gas and stellar kinematics. A
scenario in which dust is produced internally is less likely to
produce kinematic misalignments, particularly where dust originates
directly from AGB stars. Galaxy mergers in simulations often produce
misalignments \citep[e.g.][Bassett et al. submitted]{balcells90,thakar97,bendo00,dimatteo07}, thus mergers represent
a natural source for externally produced dust in ETGs. 

In this work we examine the
origins of dust in ETGs using data from the SAMI Galaxy Survey \citep{bryant15}. The majority
of SAMI galaxies are selected from the Galaxy And Mass Assembly
survey \citep[GAMA][]{driver11}, therefore we focus on the samples of
dusty ETGs selected from GAMA by \citet[][A13 hereafter]{agius13} and
\citet[][A15 hereafter]{agius15}. We begin by considering those 540 GAMA
galaxies observed by the SAMI survey that are found to have high
quality kinematic measurements (see Section
\ref{section:kinemeas}) and clearly defined visual morphologies. We
choose to explore the kinematics of A13/A15 galaxies rather than other
samples of dusty ETGs \citep[e.g.][]{rowlands12,dariush16} as we find the
largest overlap with this sample, which amounts to 49 \textit{Herschel} detected and
99 non-detected galaxies.
Together \citet{rowlands12} and \citet{dariush16} have a total of 4
galaxies currently observed by SAMI. 

This paper is structured as follows: in Sections \ref{section:samples}
and \ref{section:data} we present the samples and data-sets considered.
Section \ref{section:kinemeas} presents our method of extracting
integrated kinematic quantities from SAMI IFS observations as well as
our kinematic criteria for isolatibng those visually-classified dusty
ETGs 
that are most likely to host hot X-ray emitting gas. In Section \ref{section:results} we apply this selection
to those galaxies from A13/A15 observed by SAMI. In Section \ref{section:discussion} we discuss the
evolutionary implications of our results, and in Section
\ref{section:conclusions} we summarise
our conclusions. Throughout this work we adopt a $\Lambda$CDM
cosmology with $\Omega_{m}$ = 0.3, $\Omega_{\Lambda}$ = 0.7, and
$H_{0}$ = 70 km s$^{-1}$ Mpc$^{-1}$.

\section{Samples}\label{section:samples}

\subsection{Dusty Early Type Galaxies: \citet{agius13} and \citet{agius15}}

The parent sample of Herschel ATLAS \citep[H-ATLAS][]{eales10} detected ETGs were first
identified and analysed by A13. Briefly, H-ATLAS is a 550 square degree IR survey
using the PACS and SPIRE instruments (targeting 100-500$\mu$m) on the
Herschel space observatory with an expected detection of $\sim$250,000 galaxies.
A13 began by investigating the H-ATLAS detections for a sample
of galaxies identified as ETGs in the GAMA dataset through visual classification
\citep[][ see also Section \ref{section:morph}]{kelvin14}, with active
galaxies excluded based on the prescription of
\citet{kauffmann03a}. The sample of A13 is restricted to the redshift
range 0.013 < $z$ < 0.06 and absolute $r$-band magnitudes brighter
than $M_{r}$ = -17.4 providing a volume-limited sample in the
$r$-band. They find an H-ATLAS
detection rate of 29\% (220/771), i.e. 29\% of the visually classified ETGs in
GAMA have IR detections greater than 5$\sigma$. \citet{rigby11} show
that in the H-ATLAS science demonstration phase that their survey data
has a catalogue number density completeness of > 80\% with the
remaining 20\% missing due to noise and/or blending of sources. The
completeness for A13/A15 galaxies should be similar to this. 
Among H-ATLAS detected ETGs there is a
trend for the ratio of dust mass to stellar mass to increase for bluer
NUV - $r$ colour, implying that recent star formation is likely
associated with an increased presence of dust. 

\subsection{SAMI Overlap With A13/A15}\label{section:samiva13}

\begin{figure}
  \includegraphics[width=\columnwidth]{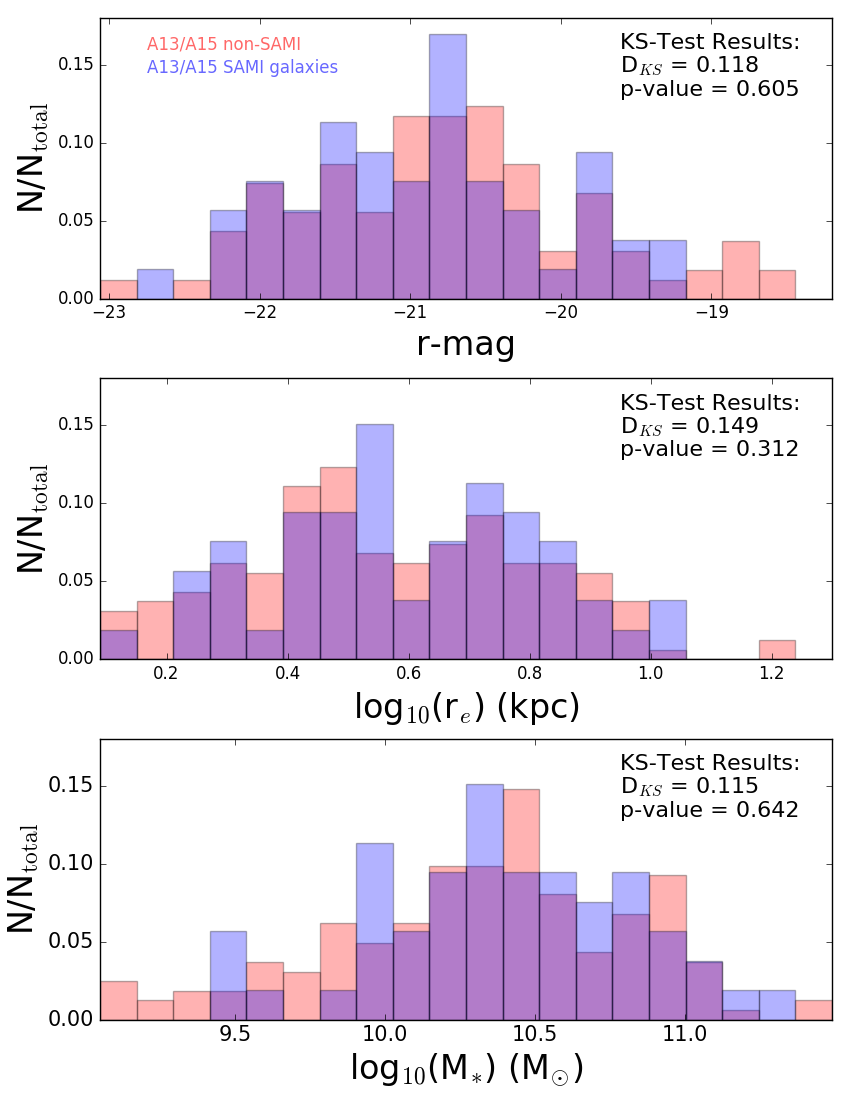}
  \caption{Histograms of $r$-band absolute magnitude,
    log$_{10}$($r_{e}$), and log$_{10}$(M$_{*}$) comparing the full
    A13/A15 H-ATLAS detected sample (red) with the subsample of
    these galaxies observed with SAMI (blue). For each galaxy
  property we perform a two sample KS-test and the results are
  indicated in the bottom right of each panel. For each of the three
  properties considered here, the KS-test results show that we can not
  reject the null hypothesis that these galaxies are selected from the
  same parent distribution.}\label{figure:sampcomp}
\end{figure}

In this paper we wish to explore the resolved kinematics of the sample
of dusty ETGs presented in A13 and
A15. While A13 includes dust masses for 220 dusty ETGs, only 49 of these
have high quality observations in the SAMI galaxy survey. Similarly,
the study of A13/A15 incudes 551 H-ATLAS non-detected galaxies, of
which 99 have high quality SAMI survey observations. We expect the
properties of those
galaxies from A13/A15 that overlap with our SAMI observations to be
fairly representative of the full sample of A13/A15 galaxies as the
SAMI survey is selected to be representative of the GAMA survey, the
parent sample of A13/A15. 

It is important to know whether the A13/A15 galaxies for which we can
investigate the resolved kinematics are representative of the original
parent distribution. In Figure \ref{figure:sampcomp} we show
histograms of $r$-band magnitude, log$_{10}$($r_{e}$), and
log$_{10}$(M$_{*}$) comparing those galaxies from A13/A15 that
have been observed with SAMI (in blue) with those that have not (in
red). We then perform a two sample Kolmogorov-Smirnov (KS) test for
each of the three galaxy properties for these two
subsamples. The resulting p-values of this are given in the top right corner of each
panel. The p-value indicates the percentage of the time we should
expect to find the observed level of difference between the two
samples, given their sizes, under the null hypothesis that they are
randomly drawn from the same parent sample (typically the null
hypothesis is not rejected where the p-value is larger than 0.01). For $r$-band magnitude, $r_{e}$, and M$_{*}$ we find p-values
of 0.605, 0.312, and 0.642 meaning we can not reject the null 
hypothesis that these samples come from the same parent
distribution. KS-test results for H-ATLAS non-detected A13/A15
galaxies show less agreement in properties between the full
sample and those observed with SAMI with p-values for $r$-band magnitude, $r_{e}$, and M$_{*}$ 
of 0.374, 0.075, and 0.132, respectively. Although we find slightly
less agreement for non-detections, the p-values suggest that, again, we can
not reject the null hypothesis that those observed by SAMI are
representative of the parent sample. 

\section{Data}\label{section:data}

\subsection{SAMI Survey Data}\label{section:samidata}

Data analysed in this work comes from the SAMI Galaxy Survey \citep{bryant15}
which aims to observe $\sim$3600 galaxies using the SAMI integral
field spectrograph \citep{croom12} at the 3.9m Anglo-Australian
Telescope in the redshift range 0.004 < $z$ < 0.095. 
Observations using the SAMI IFS represent a step forward from more
traditional IFS instruments due to the use of multiple fibre bundles
\citep[hexabundles,][]{blandhawthorn11,bryant14} allowing for
simultaneous observations of multiple galaxies with
a roughly circular, $\sim$14$\farcs$7 diameter coverage. Fibres are fed into the AAOmega spectrograph
\citep{sharp06}, which observes two spectral ranges using a red and
blue arm setup. This provides coverage with 3700-5700 {\AA} at
R=1812 resolution and with 6300-7400 {\AA} at R=4263
resolution \citep{croom12,bryant15}.

The SAMI designed to be representative of
the highly complete \citep[>98\%][]{driver11} GAMA survey rather than
complete itself due to observational constraints. As we have
mentioned, H-ATLAS detected A13/A15 galaxies should have a
completeness of >80\% similar to the overall H-ATLAS survey. We have
shown in Figure \ref{figure:sampcomp} that H-ATLAS detected A13/A15
galaxies with reliable SAMI kinematics are representative of overall
A13/A15 sample, thus we do not expect completeness issues in the SAMI
survey to affect our results.

At the time this paper was written, 1094 galaxies have been observed
by the SAMI galaxy survey. Of these, 753 have had stellar kinematics
measurements performed as described in Section
\ref{section:skinemeas}. We then perform two quality cuts on this
sample of 753 galaxies. First we utilise only those galaxy
observations that include
enough high signal-to-noise spaxels such that we can measure the
rotation curve beyond its turnover radius (see Section
\ref{section:kinemeas} for more information). Next we remove galaxies
that exhibit highly uncertain visual classifications of their
morphologies \citep[see][]{cortese16}. Our cut on stellar kinematics
quality removes 199 galaxies while the morphological cut removes a
further 14 galaxies, resulting in a final sample of SAMI survey
galaxies of 540. We use the
kinematic measurements of this large sample of galaxies to determine if a given galaxy is
supported by rotation or by
random motions. We note that our stellar kinematic quality cut removes 3 H-ATLAS
detected galaxies from A13/A15 due to large
stellar velocity dispersion errors and our morphological quality cut
removes a further 1 A13/A15 dusty ETG. All four of these galaxies,
however, have relatively low velocity dispersions, thus excluding them
does not affect our conclusions. 

\subsubsection{Stellar and Ionised Gas Kinematics}\label{section:skinemeas}

Here we briefly describe the stellar kinematics fitting process,
however, for a more detailed description see \citet{fogarty15} and van
de Sande et al. (accepted for publication in ApJ). 
Stellar kinematics are measured using the penalised pixel-fitting
\citep[pPXF,][]{cappe04} routine, which has become the standard method
for use with IFS datacubes
\citep[e.g.][]{emsellem07,jimmy13,bassett14,ma14}. The pPXF method
convolves spectral templates with a line-of-sight velocity distribution (LOSVD)
parameterised using Gauss-Hermite polynomials. The first and second
moments of this LOSVD provide the stellar velocity and velocity
dispersion, respectively. In this work we are concerned only with
these first two moments, however see \citet{vandesande17} for a detailed analysis of higher order moments. 

Ionised gas kinematics for SAMI galaxies are measured from emission
line spectra using the LZIFU
spectral fitting pipeline \citep{ho16}. Prior to fitting,
the best fitting stellar continuum model from our pPXF procedure is
subtracted from the spectra in each spaxel to provide more reliable
fits. Gas velocities and velocity dispersions are then extracted
from Gaussian fits to ionised gas emission lines. The LZIFU pipeline
provides single Gaussian fits as well as more
complex fits employing two and three Gaussian components. For the
analysis presented in this paper we are primarily interested in
the circular velocity, $V_{c}$, and kinematic position angle (both described in Section
\ref{section:kineclass}) for the
primary component of the ionised gas. Therefore we use the simple, single-component
fits. For more detail on SAMI ionised gas kinematics fits see
\citet{ho14}, \citet{ho16b}, and \citet{ho16}. 

\subsubsection{Galaxy Morphology}\label{section:morph}

Galaxy morphologies have been determined through visual classification by an
internal SAMI working group based on Sloan Digital Sky Survey
\citep[SDSS,][]{york00} Data Release 9 (DR9) RGB images for all SAMI
galaxies observed at the time this paper was written. This
classification, described in \citet{cortese16}, is independent of, but similar to, the scheme of
\citet{kelvin14} used by the GAMA survey. This involves a
step-by-step procedure in which galaxies are first broadly classified
as spheroid-dominated or disk-dominated, then placed into subclasses
based on finer details. As SAMI galaxies in this work are selected
from the GAMA survey, all galaxies here have been classified by both
teams with three key differences. First, GAMA team members performed
classifications on false colour $g$, $i$, $H$ band composite images,
while SAMI team members utilise SDSS DR9 $gri$ images. Second, the
classification working groups from both surveys are composed of
independent groups of classifiers who will each have their own unique
classification bias. Third, SAMI classifications include two criteria
for identifying LTGs not used by \citet{kelvin14}, namely the presence
of spiral arms and signs of star formation (based on colour rather
than purely on morphology). Discussion of differences between the
classifications of the two groups can be found in Section
\ref{section:classdiff}. As noted previously, 14 galaxies determined to be
``unclassified'' \citep[see][]{cortese16} are excluded from our
analysis, and only one of these comes from the A13/A15 samples. This
galaxy, although it has an H-ATLAS detection, has a low velocity
dispersion. Thus excluding it does
not affect our conclusions.
For consistency with A13/A15, galaxies with elliptical, S0, and Sa
visual classifications are defined as ETGs. 

\subsection{Stellar Mass, Dust Mass, and SFR}\label{section:dustmass}

In Section \ref{section:dustprops} we explore the dust mass scaling
relations of A13/A15 galaxies considered in this work.
The GAMA survey provides a number of ancilliary data
products, which are available for all A13/A15 galaxies observed by SAMI. Briefly, the GAMA
survey is a multiwavelength survey of hundreds of thousands of low
redshift galaxies. The core of the GAMA survey is a spectroscopic
survey at optical wavelengths using the AAOmega instrument at the
Anglo-Australian Telescope. This spectroscopic campaign is bolstered
by data sharing agreements and coordination with other independent
imaging surveys covering the entire electromagnetic spectrum, from
X-rays to radio. For more information on the goals, target selection,
and public data releases of GAMA survey data see \citet{driver09},
\citet{baldry10}, \citet{driver11}, and \citet{liske15}.

Using data products from the GAMA survey we
explore stellar masses ($M_{*}$), dust masses ($M_{d}$), and SFRs
derived from full spectral energy distribution (SED) fits to ultraviolet (UV) to far-infrared
(far-IR) observations using the \texttt{MAGPHYS} code based on the
models of \citet{dacunha08}. \texttt{MAGPHYS} has the distinct advantage over more
traditional SED fitting techniques \citep[e.g.][]{bruzual03} as the inclusion of
far-IR wavelengths allows for a direct balancing of energy from young,
hot stars and warm/cold dust emission resulting in more robust SFRs as
well as estimates of $M_{d}$. Preliminary results from 
\texttt{MAGPHYS}-determined values for GAMA galaxies have been explored by
\citet{davies16a} and \citet{driver16}, and full details of the
\texttt{MAGPHYS} analysis will be presented in Driver et al. (2017). 

Although all SAMI galaxies (including both H-ATLAS detected and
non-detected galaxies from A13/A15)
have \texttt{MAGPHYS} estimates of $M_{d}$, estimates for H-ATLAS
non-detected galaxies are highly uncertain due to the lack of far-IR
data. For this reason, we estimate
upper limits to the dust masses for H-ATLAS non-detected galaxies
following the procedure of A13/A15. This
procedure is described in Appendix \ref{appendix:dustul}. Both $M_{*}$
and SFR can more reliably be extracted in the absence of far-IR
detections, thus these values are taken from the GAMA survey for both
H-ATLAS detected and non-detected A13/A15 galaxies.

\section{Global Kinematics and Kinematic Galaxy Selection}\label{section:kinemeas}

Here we first
describe our methods of extracting the global \textbf{stellar} kinematic quantities of
rotational velocity, $V_{c}$, and flux weighted velocity dispersion,
$\sigma_{mean}$, from our IFS observations. This is followed by a
description of our method of selecting galaxies with stellar
kinematics dominated by random motions.

\subsection{Circular Velocity: $V_{c}$}\label{section:vc}

The first step in determining the stellar $V_{c}$ for each galaxy is to determine
the kinematic position angle (PA) based on the observed stellar velocity
map. This is achieved using the code \texttt{fit\_kinematic\_PA}
\citep[see e.g.][]{cappe11} on the SAMI stellar velocity maps. This
code determines the global kinematic position angle following the
method described in Appendix C of \citet{krajnovic06}.
Next, we use the measured stellar kinematic PA of each galaxy to extract the projected
rotation curves along the kinematic major axis and, from this fit,
estimate the value of $V_{c}$ using a custom Python code. We briefly
outline this procedure here, however a more detailed description is
given in Appendix \ref{appendix:vc}.

To recover $V_{c}$ we trace an artificial slit of width 1$\farcs$5
across the velocity map at an angle given by the PA. The velocity as a
function of position along the slit is fit by a piecewise
function made up of two constant velocity sections separated by a
sloped linear
segment describing the central velocity gradient. This functional
form, which follows \citet{epinat09},
provides two parameters: the turnover radius, $r_{t}$, and
$V_{c}$. The latter is given by the constant velocity value beyond
$r_{t}$. For some observations the coverage of the SAMI bundle does
not extend beyond $r_{t}$, thus measured values of $V_{c}$ are largely
unconstrained. This is true for 199 of the 753 galaxies tested
(including 3 H-ATLAS detected A13/A15 galaxies), and
these galaxies are excluded from further analysis. Finally we apply an
inclination correction to $V_{c}$ based on measured ellipticities and
bulge-to-total ratios taken from the GAMA survey and from
\citet{simard11}, respectively (our inclination correction is described fully in Appendix
\ref{appendix:vc}). 

\subsection{Flux Weighted Velocity Dispersion: $\sigma_{mean}$}

We adopt the value $\sigma_{mean}$, the flux weighted stellar velocity
dispersion, for our global velocity dispersion measure following
previous IFS studies at various redshifts
\citep{law09,epinat09,jones10,wisnioski11,green14}. 
Prior to measuring the stellar $\sigma_{mean}$, we mask spaxels with
large uncertainties on $\sigma$ following the
procedure of \citet{vandesande17}. $\sigma_{mean}$ is then defined as:
\begin{equation}
  \sigma_{mean} = \frac{\sum_{i} \sum_{j} F(i,j) \times \sigma(i,j)}{\sum_{i} \sum_{j} F(i,j)}
\end{equation}
where $F(i,j)$ is the flux observed in the spaxel with $i$ and $j$ as
its spatial position, and $\sigma(i,j)$ is the corresponding stellar
velocity dispersion. We find this measurement is robust for all galaxies with SAMI coverage
beyond $r_{t}$ as described in Section \ref{section:vc}. 
A rough correction for the effects of beam smearing is applied
following \citep{bassett14},
where the artificial $\sigma$ induced by the seeing is subtracted from
$\sigma_{mean}$ in quadrature. For a detailed description of our
masking and beam smearing correction procedure, see Appendix
\ref{appendix:bs}. For simplicity, all remaining references to $V_{c}$
and $\sigma_{mean}$
in this paper refer specifically to inclination corrected and
beam-smearing corrected values respectively. 

\subsection{Kinematic Galaxy Selection}\label{section:kineclass}

In this
Section we utilise galaxy stellar kinematics from IFS
observations to select dispersion-dominated galaxies in a less
ambiguous way than visual morphological classification. 
We would like to know how many galaxies that are visually classified as ETGs
are really
dispersion-dominated systems, and how many have kinematic properties 
more similar to rotationally-supported LTGs. The latter are typified
by S0 galaxies, which, by definition, exhibit a significant disk
component.

\begin{figure}
  \includegraphics[width=\columnwidth]{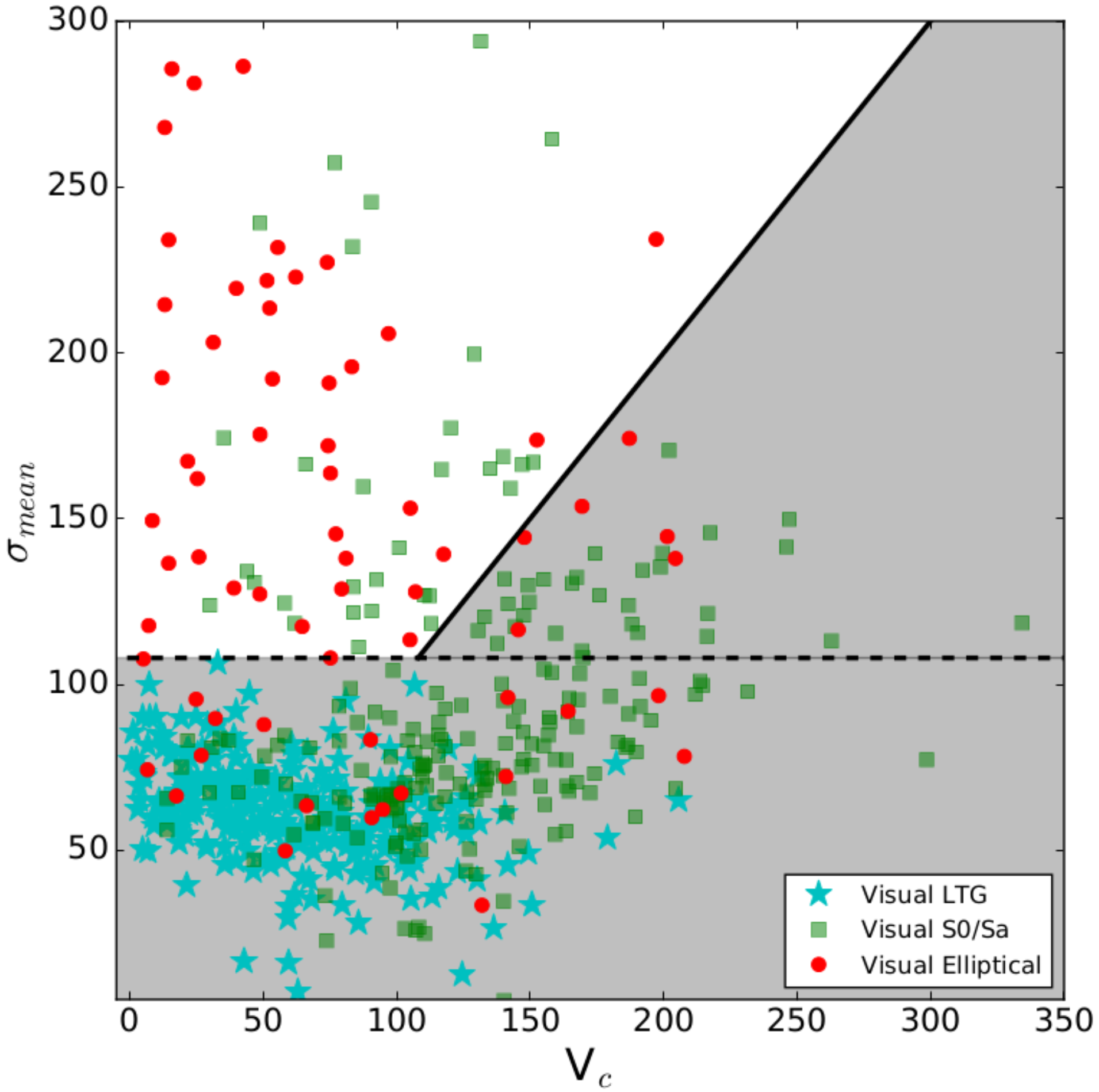}
  \caption{A depiction of our kinematic classification scheme based on
  $\sigma_{mean}$ versus $V_{c}$. We compare $V_{c}$
  versus $\sigma_{mean}$ for three subsets in visual morphology: galaxies
  classified as elliptical by both SAMI and GAMA groups in (red circles),
  galaxies classified by both groups as later than Sa (blue stars), and galaxies
  classified as S0 or Sa by the SAMI team (green squares). The solid line black shows the 1-to-1
  relation, the black dashed line shows our $\sigma_{mean}$ cutoff
  isolating visual LTGs. The black solid and black dashed lines are used to separate our
  three kinematic selections.}\label{figure:kcl}
\end{figure}

Our stellar kinematic selection is depicted in Figure \ref{figure:kcl}
where we plot $\sigma_{mean}$ versus $V_{c}$. Plotted symbols indicate
different visual morphologies taken from our SAMI
classifications.  We note that the
velocity resolution of the SAMI survey is 70 km s$^{-1}$, which means
that many of our low stellar $\sigma_{mean}$ values will
be upper limits. In particular, this will be the case for a very large fraction of
visually classified LTGs at low $V_{c}$, as this low measurement of
$V_{c}$ often results from a nearly face-on inclination. Measurements
for face-on galaxies should provide a lower value of $\sigma$ when
compared to an edge-on view of the same object due to a minimal
contribution from rotation and beam smearing.  In their IFS study
of face-on LTGs from the DiskMass Survey, \citet{martin13}
find that 77\% (23/30) have line-of-sight stellar $\sigma$ less than
70 km s$^{-1}$ with an average value of 56.8 km s$^{-1}$ for their
entire sample. Thus we should expect low $V_{c}$ (more face-on on
average) galaxies have $\sigma_{mean}$ clustered near our $\sigma$
resolution limit. We also note that we apply a larger
beam smearing correction for galaxies with a large $V_{c}$ (see
Appendix \ref{appendix:bs}), and the uncorrected measurements of
$\sigma_{mean}$ for these galaxies are up to 30 km s$^{-1}$ larger
than pictured in Figure \ref{figure:kcl}.  

Visually classified ETGs (Elliptical, S0, and Sa galaxies) are found to
exhibit a large amount of scatter in $\sigma_{mean}$ in Figure
\ref{figure:kcl} highlighting the pitfalls of assuming a one-to-one
correspondence between visual-morphology and kinematics;
e.g. visually-classified ETGs have a large stellar velocity dispersion.
Visually classified LTGs, on the other hand, are found to be more clustered. This is due to
the fact that they are easier to identify from the 
presence of clear spiral arms, resulting in a much cleaner
selection. S0/Sa galaxies, in general, extend the high $V_{c}$ end of the LTG
distribution to higher $\sigma_{mean}$. This is consistent with the
result of \citet{williams10} who show S0 galaxies exhibit a larger
$V_{c}$ than LTGs at fixed M$_{*}$. 

Following the LTGs, we produce a selection
to separate galaxies having kinematic properties consistent with those
of visually selected LTGs. We initially perform a linear fit to the
visual LTGs in Figure \ref{figure:kcl}, finding a slope of -0.04$\pm$.04.
As this is consistent within errors to a flat slope, we simply employ a flat cut in
$\sigma_{mean}$, matching the cutoff value to the highest value
observed for a visually selected LTG of 108.0 km s$^{-1}$. This cut is
shown in Figure \ref{figure:kcl} by th horizontal, black dashed line.
By design, this isolates 100\% of LTGs
in our sample, however 31\% of visually-classified elliptical galaxies
also fall below this line (20/65). 

We are interested in galaxies for which a large fraction of the
dynamical support comes from random motions, implying a large
$\sigma_{mean}$ relative to $V_{c}$. For this reason we also plot in both
panels the 1-to-1 relation as a solid black line. Galaxies
falling above this line have $\sigma_{mean}$ > $V_{c}$, thus they are
the most likely to derive a majority of their support from random motions
\citep[e.g.][]{weiner06,law09,lemoine10,newman12}. We define galaxies
falling above both this line and the black dashed line as ``dispersion-dominated'' galaxies. The
remaining galaxies we define as ``rotation-dominated'' galaxies. Comparing this kinematic
selection with our sample of 540 SAMI galaxies with reliable kinematics,
we find that
100\% of dispersion-dominated galaxies (DDGs) and 39\% (181/469) of
rotation-dominated galaxies (RDGs) are visually classified as ETGs. We
reiterate the point from Section \ref{section:morph} that our
definition of ETG includes Sa galaxies, however. If we redefine ETGs
more strictly as only galaxies with visual classifications earlier
than Sa, we find 93\% (66/71) of DDGs and 14\% (66/469) of RDGs are
considered ETGs. 

\begin{figure}
  \includegraphics[width=\columnwidth]{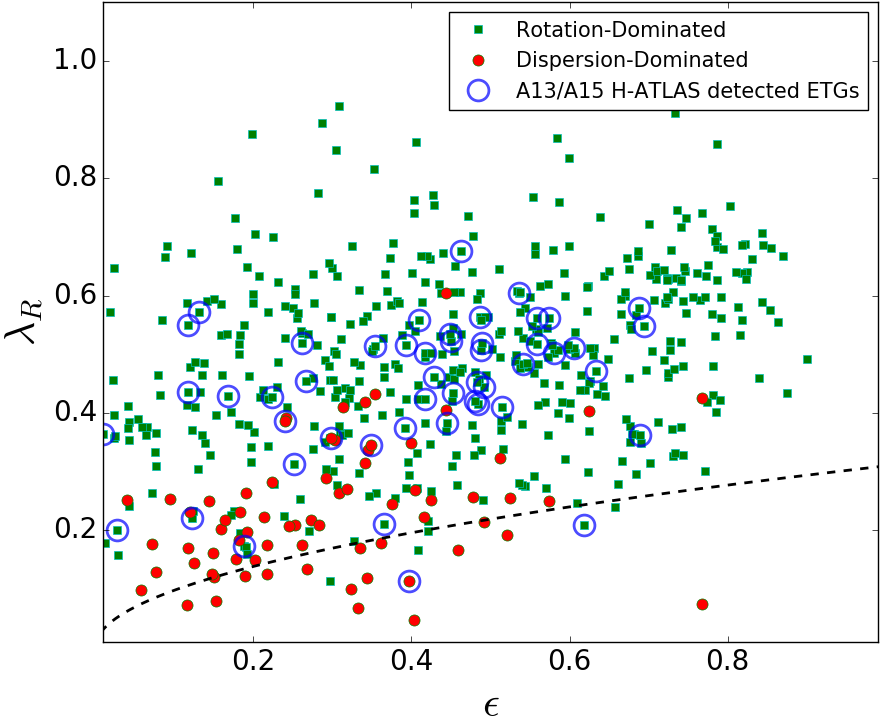}
  \caption{$\lambda_{R}$ versus $\epsilon$ for SAMI galaxies, with
    A13/A15 H-ATLAS detections highlighted. Here we show SAMI
    rotation-dominated and dispersion-dominated galaxies with cyan
    and green squares, respectively. The separation between slow and fast rotators from
    \citet{emsellem11} for ATLAS$^{\textrm{3D}}$ galaxies is shown
    with a black dashed line. We find that only two H-ATLAS detections are
    slow rotators and one exhibits rapid rotation based on our
    kinematic selection.}\label{figure:elllr}
\end{figure}

Before moving on, we examine
the relationship between the galaxy spin parameter, $\lambda_{R}$, and
ellipticity, $\epsilon$, for A13/A15 galaxies. Here, $\lambda_{R}$ is
calculated from our SAMI stellar kinematics maps as:
\begin{equation}\label{equation:lamr}
  \lambda_{R} = \frac{\sum\limits_{k=1}^n F_{k}R_{k}|
    V_{k}|}{\sum\limits_{k=1}^n F_{k}R_{k}\sqrt{V_{k}^{2}+\sigma_{k}^{2}}}
\end{equation}
Where $F_{k}$ is the flux in spaxel $k$, and $V_{k}$ and $\sigma_{k}$ are the
line-of-sight velocity and velocity dispersion in spaxel $k$. The
value $R_{k}$ is the semimajor axis of the ellipse defined by the
$r$-band axis ratio ($b/a$) on which spaxel $k$ lies (i.e. the intrinsic
radius). This sum is performed using only spaxels within an ellipse
defined by the galaxy effective radius, $R_{e}$, and $b/a$. For
ATLAS$^{\textrm{3D}}$ galaxies, \citet{emsellem11} show that these
parameters are useful in separating fast and slow-rotators among their
sample of ETGs. $\lambda_{R}$ versus $\epsilon$ for our sample is shown in Figure \ref{figure:elllr} where we
plot RDGs, DDGs, and H-ATLAS detected ETGs from A13/A15. The dashed
line shows the separation between slow- versus fast-rotators taken
from \citet{emsellem11}. 

We find that there is a correspondence between $\lambda_{R}$ vs
$\epsilon$ and our kinematic selection, with a majority of DDGs
falling at low $\lambda_{R}$ and $\epsilon$. This suggests that the
two methods are tracing similar properties of SAMI galaxies,
particularly in light of the large uncertainty in $\lambda_{R}$ for
SAMI observations (which in some cases is >0.4). Considering only H-ATLAS
detected ETGs from A13/A15, we find that employing the slow-
versus fast-rotator selection of \citet{emsellem11} would retain only
two galaxies, with one having significant rotation and a relatively
large $\epsilon$. We are able to double our sample of H-ATLAS
detected ETGs by employing the kinematic selection outlined
here. We stress that, within uncertainties in our kinematic
measurements, our kinematic selection and that of \citet{emsellem11}
are tracing roughly the same population. In this work, however, we are
primarily interested in galaxies with a large
stellar velocity dispersion, which can be used as an indication of the
presence of a hot, X-ray emitting halo
\citep[e.g.][]{boroson11,sarzi13,goulding16}. 

Although stellar
$\sigma$ is often used
as a proxy for $M_{*}$ \citep[e.g.][]{faber76}, it has been
shown that even massive, X-ray halo hosting galaxies can host disks of
cold gas and dust when rotating rapidly
\citep{negri14a,negri14b}. Using our kinematic
quantities, however, we can identify those galaxies likely to host hot X-ray
emitting gas, and further select only those low rotation galaxies in
which the presence of this gas would hinder the formation of
long-lived dust grains. Figures \ref{figure:kcl} and
\ref{figure:elllr} show that this may not be accomplished considering $\lambda_{R}$
versus $\epsilon$ or by using $M_{*}$ alone as an indicator of a hot
interstellar medium. 

\section{Results}\label{section:results}

\subsection{Kinematics of A13/A15 Galaxies}

Having developed a stellar kinematic selection, we now
apply this to those A13/A15 galaxies that have
been observed by the SAMI Galaxy Survey. The $V_{c}$ versus
$\sigma_{m}$ parameter space used to perform our kinematic selection
is depicted in Figure \ref{figure:agiuskcl} for A13/A15 galaxies
observed by SAMI with H-ATLAS non-detections shown by small cyan
circles and H-ATLAS detected galaxies shown by larger red circles. We
also indicate those galaxies having kinematic irregularities
(described below) by green
squares and blue pentagons.

From Figure \ref{figure:agiuskcl} it can be seen that a far larger
fraction of DDGs are non-detections in the H-ATLAS survey. Indeed,
considering all DDGs from A13/A15 11\% (4/35) are H-ATLAS detections compared
with 40\% (45/113) of RDGs. The entire sample of A13/A15 represents
771 galaxies with 220 of these being H-ATLAS detections, or
29\%. This clearly shows that, although a low fraction of visually
classified galaxies host appreciable amounts of dust, it is far more
likely for galaxies with kinematics dominated by rotation. In the
following Sections, we examine more closely the kinematics of H-ATLAS
detected and non-detected galaxies from A13/A15. 

\subsubsection{Kinematics of H-ATLAS Detected ETGs}\label{section:detgkine}

\begin{figure}
  \centering
  \includegraphics[width=\columnwidth]{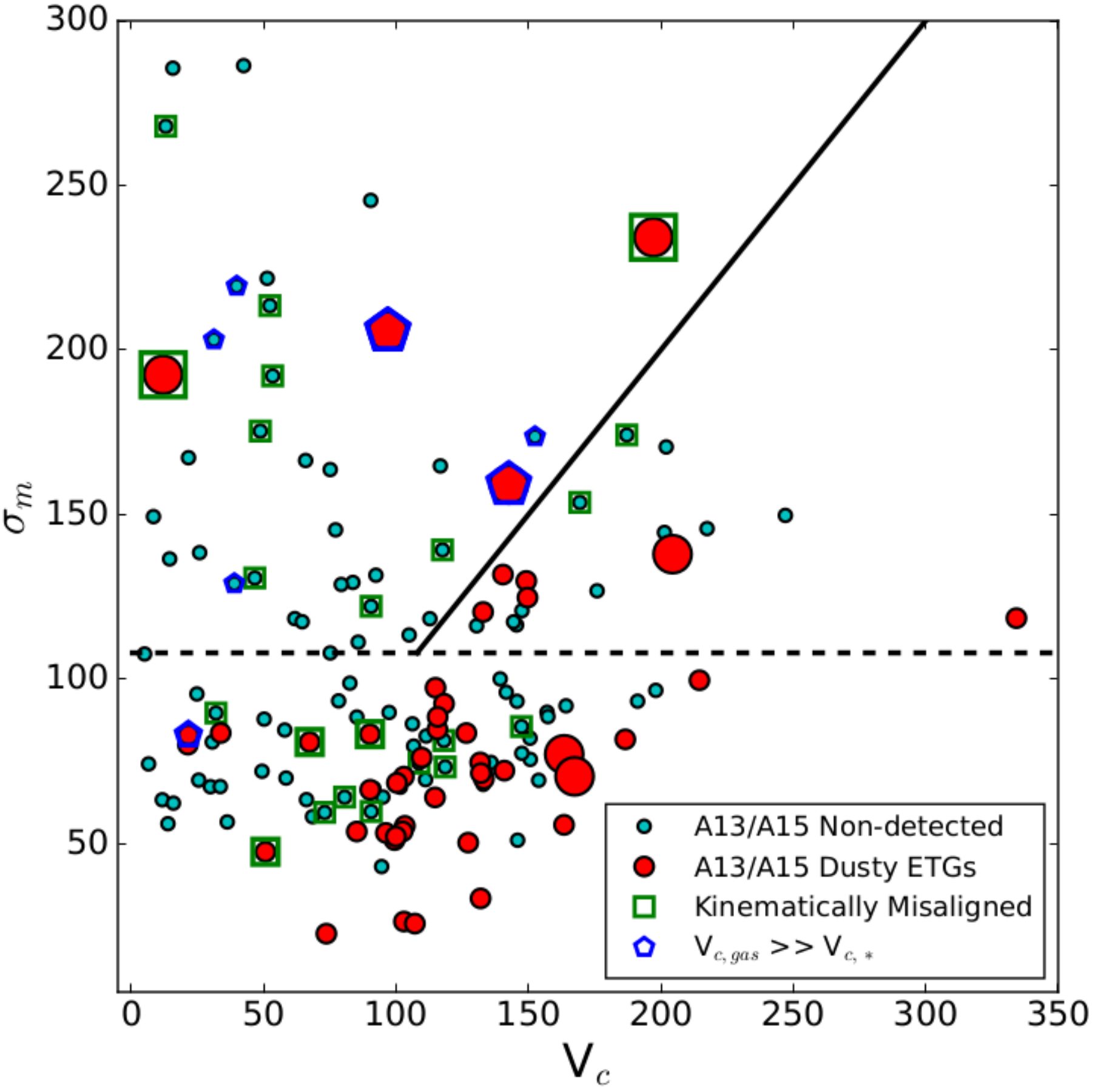}
  \caption{Stellar $V_{c}$ vs $\sigma_{mean}$ for H-ATLAS detected and
    non-detected ETGs from
    A13/A15. H-ATLAS non-detected galaxies are shown by the
    smaller cyan circles while H-ATLAS detected galaxies
    are indicated by the larger red circles. Red circles that are
    plotted with significantly larger symbols show those galaxies with
    log$_{10}$($M_{*}$) > 10.8, which are most likely to host a hot
    X-ray halo (see Section \ref{section:dustprops}). Note however that some
    H-ATLAS non-detected galaxies have log$_{10}$($M_{*}$) > 10.8, but
    these are not plotted with larger simbols. We also
    show kinematically misaligned galaxies with green squares, and
    galaxies for which the ionised gas rotation is significantly larger
    than that of the stars with blue pentagons.}\label{figure:agiuskcl}
\end{figure}

As mentioned in Section
\ref{section:samiva13}, only 49 of the 220 H-ATLAS detected ETGs of
A13/A15 have kinematics maps from the SAMI survey that meet our quality
cuts, and this
subset is shown in our $\sigma_{mean}$ vs $V_{c}$ diagram in Figure
\ref{figure:agiuskcl} with red circles. We find that 45/49 (90\%) are RDGs, with 35 of these 45 having
$V_{c}$ > 100 km s$^{-1}$. This means that these galaxies derive
a majority of their dynamical support from rotation as expected
for LTGs, particularly for S0/Sa galaxies (as shown in Figure
\ref{figure:kcl}). Galaxies such as these may host an X-ray emitting
halo if they are massive enough \citep[e.g.]{anderson15}, however
\citet{negri14a} and \cite{negri14b} have shown that
rapid rotation can allow a galaxy to host a cold gas disk even in the
presence of such a hot halo. Therefore the
presence of dust in these systems \textit{does
not} require an external origin scenario such as galaxy mergers.

\begin{figure*}
  \centering
  \includegraphics[width=5in]{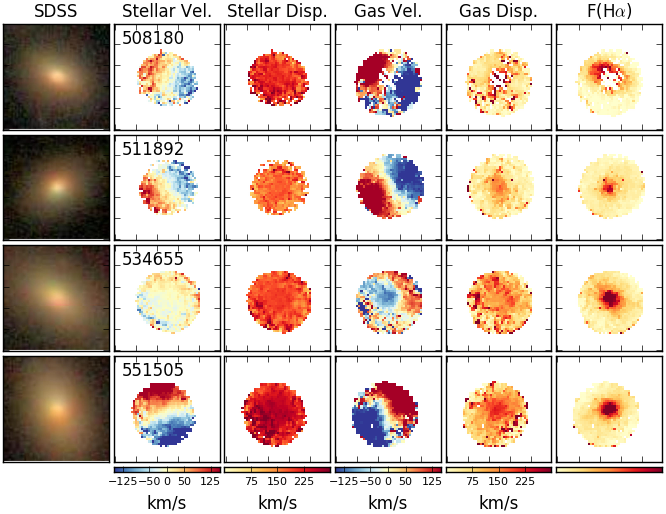}
  \caption{A montage showing morphologies and stellar and gas
    kinematics for the four H-ATLAS detected galaxies from A13/A15 identified
    as DDGs from SAMI IFS observations. From
    left to right: SDSS $gri$ image, stellar velocity
  map, stellar $\sigma$ map, ionised gas velocity map, ionised gas
  $\sigma$ map, and H$\alpha$ flux map. The size of the SDSS image is
  matched to that of the SAMI kinematics maps at
  25$\farcs$0x25$\farcs$0. Columns 2-6 are derived from
  SAMI survey observations. The GAMA CATIDs of each galaxy
are indicated above the stellar velocity maps. Colour bars for
kinematics maps are in km s$^{-1}$ while, for a given galaxy, H$\alpha$ flux maps are displayed
with a flux scale ranging from 0 to 3 times the standard deviation
of H$\alpha$ flux for that galaxy. Note that 508180 exhibits
relatively weak Balmer emission in the central regions and poorly fit
Balmer absorption, resulting in the missing pixels for gas
measurements of this galaxy.}\label{figure:agellip}
\end{figure*}

Next we investigate kinematics in our sample of four H-ATLAS
detected DDGs. A comparison between kinematics of the gas and stars
for these four galaxies can be seen in Figure
\ref{figure:agellip}, alongside their SAMI stellar and ionised gas
kinematics and their H$\alpha$ flux maps. All four galaxies exhibit
discrepancies between their stellar and ionised gas kinematics,
indicating that the dust in these galaxies is related to significant accretion
events, such as galaxy mergers, in their evolutionary histories. Here
we identify two classes of ``kinematically irregular'' galaxies,
described in the following.

The first class of kinematically irregular galaxies are
those with significant misalignments between the kinematic position
angles of the stars and ionised gas. These galaxies have been
identified by Bryant et al. (in preparation), and here we define
kinematic misalignments as those
galaxies with differences between the stellar and kinematic position
angle of >30$^{\circ}$. Kinematically misaligned galaxies such as these
show the clearest evidence from SAMI observations of having undergone
a stochastic event, such as gas accretion, in the relatively recent past
\citep[similar to non-SAMI works,
e.g.][]{knapp89,mcdermid06,davis11,davis15,vandevoort15}. Bryant et
al. (in preparation) find that the cut in stellar versus gas position
angle of 30$^{\circ}$ may not always provide an accurate descriptor of
the fraction of misaligned galaxies due to observational issues
(e.g. the depth of the data, see Bryant et al. in preperation),
however this will not affect our conclusions as kinematically
misaligned galaxies included in our H-ATLAS detected DDG sample have
misalignments close to 90$^{\circ}$. Such a large difference between
stellar and gas kinematics gives the clearest indication of recent accretion.

Galaxies 551505 and 534655 fall into this first class, with
both exhibiting kinematic misalignments of $\sim$90$^{\circ}$. The two cases are not identical,
however. 551505 displays rapid rotation in both stars and ionised gas
measured to the edge of the SAMI fibre bundle, which may suggest that
this is a polar ring galaxy, a relatively stable configuration
resulting from merger activity \citep[][Bryant et al. in
preparation]{bekki98b,iodice15}. Galaxy 534655, on the other hand, exhibits very slow
stellar rotation with a rapidly rotating ionised gas component in the
central region possibly indicative of a nuclear starburst. From
preliminary analysis of emission line ratios in the central region of
this galaxy we find possible evidence of a low-ionization nuclear
emission-line (LINER) like emission (Medling et al. in preparation), consistent with this picture. Nuclear
starburst activity such as this has also been linked to merger activity in
local luminous IR galaxies
\citep[LIRGS,][]{sanders99,bekki00,hopkins06,haan13}. 

In addition to kinematically misaligned galaxies, we also identify a
second class of galaxies
in which the stars and gas are kinematically
aligned but have a significantly larger gas $V_{c}$ when compared to
that of the stars. In order for a galaxy to be included in this
classification we require the ratio of stellar to gas rotation,
$V_{c,star}$/$V_{c,gas}$, to be <0.6 noting that values observed in
LTGs as a result of asymmetric drift \citep[a phenomenon related to
the aging of stellar populations][]{gomez77,westfall07} fall in the range $\sim$0.75-0.9
\citep{martin13,cortese14,cortese16}. If the origin of the ionised gas content of an ETG
were closely related to the existing stellar component, we would
expect the two to share similar kinematics, unlike what we see in such
cases. This implies that
$V_{c,star}$/$V_{c,gas}$ < 0.6 galaxies have experienced an accretion
event in the past related to their gas and dust content, but the
difference between the time this occurred and the time at which we
observe the galaxy may be significantly longer than for kinematically
misaligned galaxies depending on the dynamical relaxation time of the
system. Estimates of the relaxation time of gas disks in merger
remnants range from $<<$ 1 Gyr to $\sim$5 Gyr
\citep{lake83,davis16}, however see Bryant et al. (in preparation) for
discussion of dynamical relaxation time in SAMI survey galaxies.

The other two galaxies in Figure \ref{figure:agellip}, 508180 and 511892, fall into our second class
of kinematic irregularities. As noted, the typical values of
$V_{c,star}$/$V_{c,gas}$ seen in LTGs due to asymmetric drift are
$\sim$0.75-0.89, whereas in galaxies 508180 and 511892 this
value is roughly half that at 0.35 and 0.49 respectively. One
scenario would be a prograde minor merger where gas is
accreted with a similar angular momentum as the accreting
galaxy. Retrograde merger remnants are more likely to exhibit gas-stellar
counter rotation after dynamical relaxation,
particularly in cases where the primary galaxy is gas poor prior to
the merger (e.g. Bassett et al., submitted to MNRAS). 

Of the 45 H-ATLAS detected RDGs, four also show kinematic discrepancies
similar to the four dusty DDGs. This is
reasonable as minor mergers are not limited to massive,
dispersion-supported galaxies. It is important to note that, although the
dust content of some fraction of low-dispersion galaxies will indeed
be related to accretion processes, these processes are not a necessity
to account for the observed dust in the absence of a hot, X-ray
halo. We also indicate with large symbols those H-ATLAS detected ETGs
with log$_{10}$($M_{*}$) > 10.8, which \citet{anderson15} find is the
limiting mass above which galaxies show clear evidence for an X-ray
emitting ISM (see Section \ref{section:dustprops}). Three RDGs fall in this category, however they all have
$V_{c}$ > 160 km s$^{-1}$. \citet{negri14a} and \citet{negri14b} show
in simulations that, even in the presence of an X-ray halo, rapid
rotation can allow for the presence of a cold gas disk.
Furthermore, two of these massive RDGs
have $\sigma_{mean}$ $\simeq$ 75 km s$^{-1}$, suggesting that
these two galaxies do not follow the Faber-Jackson relation for
massive ETGs \citep{faber76}. This discrepancy between their large
$M_{*}$ with a low $\sigma_{mean}$ clearly illustrates that these
galaxies must derive a significant amount of support from rotation.

\subsubsection{Kinematics of H-ATLAS Non-Detected ETGs}\label{section:ndetgkine}

In this Section, we
briefly discuss the integrated kinematics of H-ATLAS
non-detections from A13/A15. These galaxies
are indicated in Figure \ref{figure:agiuskcl} as small cyan
circles. Similar to H-ATLAS detected ETGs, we
find that non-detected galaxies also
occupy the full range in $\sigma_{mean}$ versus $V_{c}$ as the 540
SAMI galaxies with reliable measurements, including a significant
number of RDGs. We do find, however, that a larger percentage of
non-detected galaxies fall in our DDG kinematic selection at 31\%
(31/99) compared to 8\% (4/49) for H-ATLAS detections.

Next we examine the level of kinematic irregularity among our 31 H-ATLAS
non-detected DDGs. As discussed in the previous section, kinematically
irregular galaxies are thus defined based on a comparison of their
stellar and ionised gas kinematics. While all four of our H-ATLAS
detected DDGs have strong ionised gas emission, only 35\% (11/31) of H-ATLAS
non-detected DDGs have ionised gas emission with a high enough
signal-to-noise to evaluate this. Thus a majority (20/31) of H-ATLAS
non-detected DDGs in our sample are poor in gas as well as dust, as is
typical of low redshift ETGs. Among the H-ATLAS non-detected DDGs with
ionised gas emission strong enough to measure rotation, 7/11 have
kinematically misaligned gas and 4/11 exhibit $V_{c,star}$/$V_{c,gas}$
< 0.6 (specifically 0.09, 0.10, 0.27, and 0.56). This is similar to
kinematic irregularities seen in H-ATLAS detected DDGs, therefore the
presence of dust does not impact the relative dynamics of gas compared
to stars in ETGs with significant gas. We note, however that the
presence of ionised gas in the
absence of a secure detection of dust emission is not inconsistent
with the presence of a hot, X-ray emitting ISM in massive ETGs. We discuss this point
further in Section \ref{section:dustdisc}. 

For completeness we note that
among the full sample of DDGs in our 
SAMI kinematics sample, 58\% (41/71) have a high ionised gas emission
with high enough signal-to-noise to measure rotation. Among this
subsample, 22/41 exhibit kinematic misalignments while 13/41 fall in
the $V_{c,star}$/$V_{c,gas}$ < 0.6 class. Given our small sample size,
our finding that 54\% of
DDGs in our sample with appreciable amounts of ionised gas are
kinametically misaligned agrees well
with the work of Bryant et al. (in preparation). The authors find that,
depending on the exact definitions, $\sim$40-53\% of ETGs from the
full SAMI survey with high
signal-to-noise ionised gas emission are kinematically misaligned.

\subsection{Dust Properties of H-ATLAS Detected ETGs}\label{section:dustprops}

\begin{figure}
  \includegraphics[width=\columnwidth]{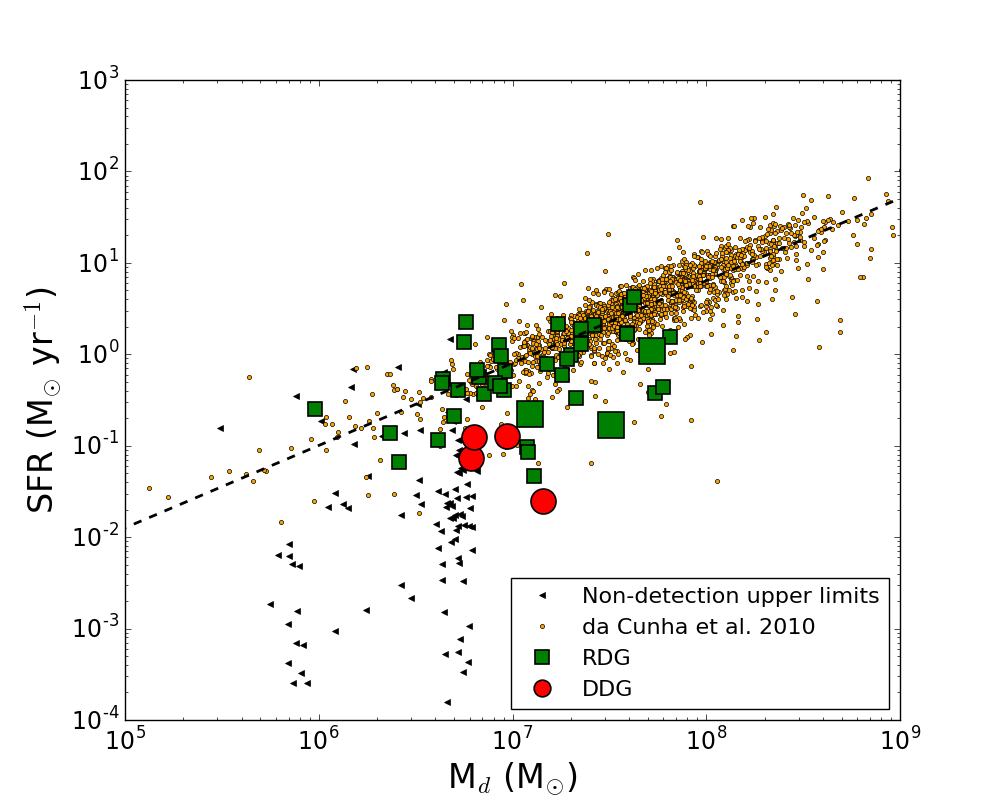}
  \caption{The relationship between $M_{d}$ and SFR for dust ETGs of
    A13/A15 that have been observed by SAMI in comparison with normal
    $z=0$ star-forming galaxies from \citet{dacunha10a}. Larger
    symbols indicate galaxies with log$_{10}$($M_{*}$) > 10.8, which
    are most likely to host a hot X-ray halo (note all DDGs here fall
    into this selection). Upper limits for
    \textit{Herschel} non-detected galaxies from A13/A15 are also indicated
    with black triangles. Normal
    star-forming SDSS galaxies from \citet{dacunha10a} follow a tight
    relationship 
    between $M_{d}$ and SFR and the linear fit to these datapoints is
    shown by the dashed black line. A majority of RDGs follow this
    relation while DDGs fall below,
    having low SFR for their dust content when
    compared with normal SFR galaxies.}\label{figure:mdsfr}
\end{figure}

\begin{figure}
  \includegraphics[width=\columnwidth]{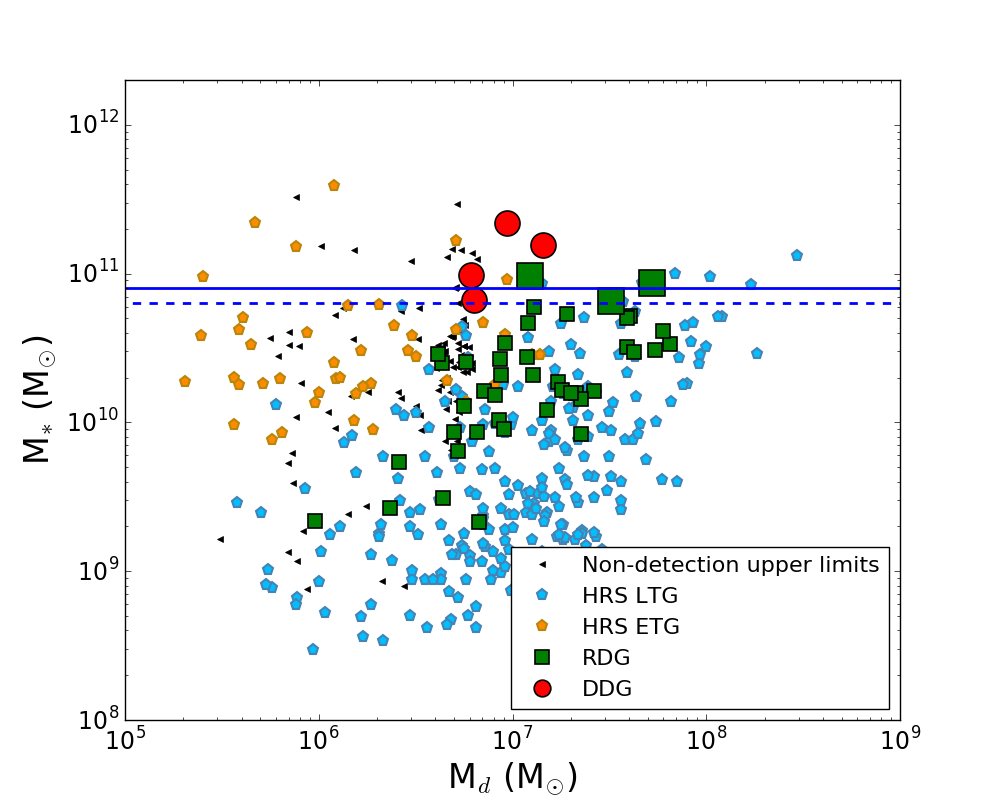}
  \caption{$M_{d}$ versus $M_{*}$ for SAMI observed A13/A15
    galaxies (symbols match those in Figure \ref{figure:mdsfr}), as well as very nearby galaxies from
    the Herschel Reference Survey \citep[HRS;][]{boselli10}. Blue and
    orange pentagons indicate HRS LTGs and HRS ETGs, respectively.
    We also show with horizontal blue lines the
    mass limit above which all elliptical galaxies from
    \citet{anderson15} show evidence of an extended X-ray emitting halo.}\label{figure:mdms}
\end{figure}

Having shown that a majority of the visually classified dusty, ETGs
from A13/A15 are consistent with being rotationally-supported, we now
investigate the dust content of our sample. Any differences between
members of our kinematic selections (or lack thereof), may help to further
identify the most likely origin scenario for their dust content. We present in Figures
\ref{figure:mdsfr} and \ref{figure:mdms} the
$M_{d}$-$SFR$ and $M_{d}$-$M_{s}$ relationships for our sample, with
markers indicating our kinematic $V_{c}$-$\sigma_{mean}$
selection. We also include upper limits on $M_{d}$ for H-ATLAS
non-detections from A13/A15 as small black triangles. 

In Figure \ref{figure:mdsfr},
our observations of $SFR$ vs $M_{d}$ for A13/A15 galaxies are plotted over a large
sample of ``normal'' star-forming SDSS galaxies taken from
\citet{dacunha10a}, which represent the $z=0$ star-forming main
sequence, plotted as small orange dots. Also plotted in Figure \ref{figure:mdsfr} is a linear fit to
$M_{d}$ vs $SFR$ from \citet{dacunha10a} given by the black dashed
line. 

In
Figure \ref{figure:mdms}, $M_{*}$ vs $M_{d}$ for A13/A15 galaxies are plotted over galaxies from the
Herschel Reference Survey \citep[HRS][]{boselli10,cortese12b},
representing a wide range of galaxy types and environments. Visually
classified LTGs from the HRS are given by light blue pentagons while orange pentagons
show ETGs. Visual classifications for HRS galaxies are more
reliable than those of SAMI galaxies, because HRS galaxies are extremely
nearby objects. The relative proximity of HRS galaxies also means that
they are sensitive to much lower levels of total $M_{d}$ than A13/A15,
which is reflected in Figure \ref{figure:mdms} where HRS ETGs overlap
with A13/A15 upper limits. This selection effect, however, will not
affect our conclusions. We also plot in Figure \ref{figure:mdms} blue horizontal lines
that correspond to stellar masses of galaxies from \citet{anderson15}
that show evidence of an extended X-ray emitting halo. The solid blue
line is located at log$_{10}$($M_{*}$/$M_{\odot}$) = 10.9, above which all
galaxies show clear evidence of such a halo. Considering tentative
detections, this can be extended to log$_{10}$($M_{*}$/$M_{\odot}$) = 10.8,
indicated by the dashed blue line. \citet{anderson15}
explore the relationship between X-ray luminosity ($L_{X}$) down to individual
galaxy masses of log$_{10}$($M_{*}$/$M_{\odot}$) = 10.0 using stacking of X-ray
observations. Below log$_{10}$($M_{*}$/$M_{\odot}$) = 10.8 they find no dependence
between $L_{X}$ and $M_{*}$, suggesting that the observed $L_{X}$ can
be explained by SNe remnants and X-ray binaries rather than a hot gas
halo. The results of \citet{anderson15} suggest that an alternative to
using kinematics to select galaxies hosting X-ray gas halos is to
employ a fixed stellar mass limit of log$_{10}$($M_{*}$/$M_{\odot}$) = 10.8.

Figure
\ref{figure:mdsfr} shows that DDGs are found to host a lower SFR at fixed $M_{d}$
when compared to the bulk of RDGs; this
difference is even greater when comparing to SDSS star-forming
galaxies \citep[a linear fit from][is shown as a black dashed
line]{dacunha10a}. In Figure \ref{figure:mdms} it can also be seen
that the DDGs are among the most massive galaxies in our sample, and
they host extremely small dust reservoirs given their stellar masses.
consistent with the assertion DDGs represent genuine
massive, elliptical galaxies, likely to host a hot, X-ray
emitting interstellar and/or intergalactic medium \citep{anderson15}. 

Irregularities between
stellar and gas kinematics favour a merger driven explanation for the
dust content of H-ATLAS detected DDGs. In
this scenario, these galaxies begin as typical quiescent ellipticals
hosting very little molecular gas and dust
\citep{leeuw08,young11,smith12}, thus occupying the upper left of
Figure \ref{figure:mdms}. These galaxies will then undergo
minor mergers
with gas rich satellites containing 
both star-forming gas and dust that is stripped by the central
galaxy. A minor merger such as this will significantly increase $M_{d}$ while
contributing negligibly to $M_{*}$, thus moving galaxies horizontally
towards the right. This is consistent with their location in Figure
\ref{figure:mdms}, offset from HRS LTGs and RDGs. Observations have
also shown that the star formation efficiency of
gas stripped from galaxies can be extremely low
\citep{knierman13,jachym14}, consistent with $M_{d}$ versus SFR for
kinematic ETGs presented here.

RDGs more closely follow the relationship for normal
star-forming galaxies of \citet{dacunha10a} in Figure
\ref{figure:mdsfr} than DDGs, and have a $M_{d}$-$M_{*}$ relationship
consistent with HRS LTGs. Although there are examples of suppressed
star-formation at a wide range of dust masses, these are found
to be within the scatter of the SDSS data. Recently \citet{lianou16}
examined the scaling relations for ETGs in the HRS finding a
significantly larger scatter for ETGs than that observed by
\citet{dacunha10a}, with galaxies typically deviating to low SFR,
consistent with results presented here. A possible explanation for the
position of low SFR RDGs in Figure \ref{figure:mdsfr} is
morphological quenching \citep{martig09}, where the efficiency of
converting molecular gas into stars is reduced in the presence of a
massive bulge. This has been seen in observations previously
\citep{saintonge12} and, given the known correlation between $M_{*}$
and bulge-to-total ratio \citep[e.g.][]{lang14}, can also explain why
all three RDGs with log$_{10}$($M_{*}$) > 10.8 $M_{\odot}$ exhibit a
low SFR with a retention of their dust content. 

\section{Discussion}\label{section:discussion}

\subsection{Moving Beyond Visual Classification of Galaxies}\label{section:classdiff}

\begin{figure}
  \centering
  \includegraphics[width=\columnwidth]{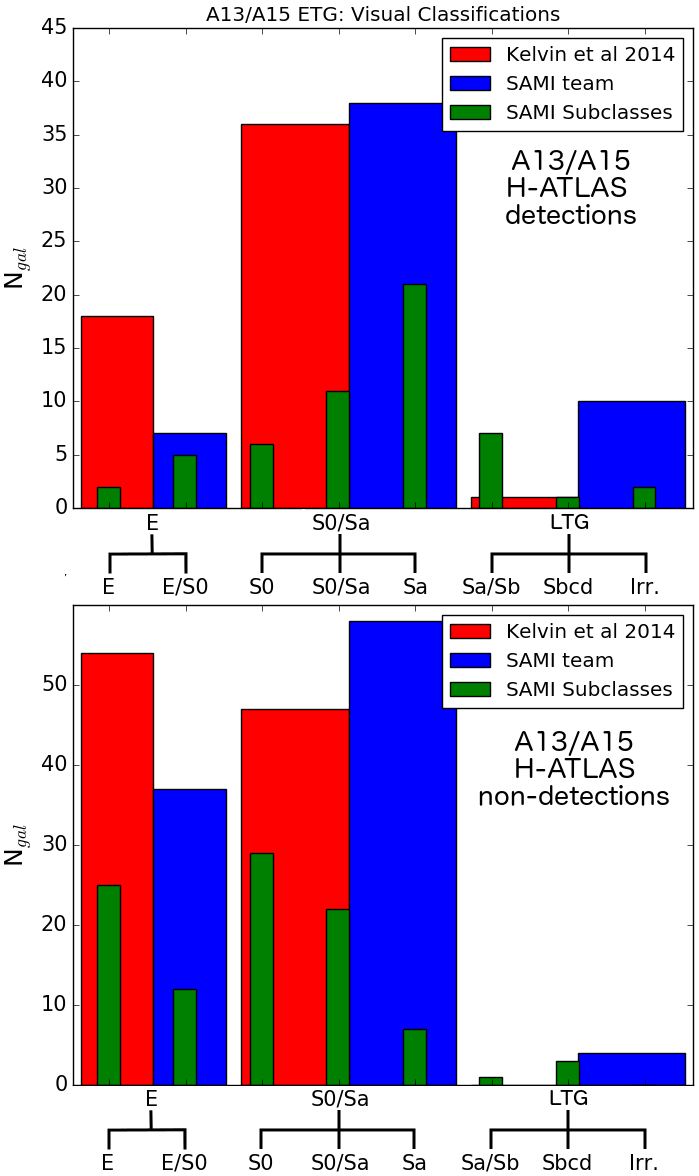}
  \caption{Comparison between visual morphologies from
    \citet{kelvin14} and those of the SAMI team. The classifications
    of \citet{kelvin14} and are based on \textit{g}, \textit{i}, and \textit{H} band
    images while those of SAMI, using a similar classification scheme,
    are based on SDSS $gri$ images from DR9. In both panels, large red
    and blue bars show broad classifications (E, S0/Sa, LTG) from the
    GAMA and SAMI teams, respectively, while small green bars show
    subclasses for the SAMI team. The upper axis labels correspond to
    the broad classes while the lower labels correspond to SAMI
    subclasses. \textbf{Top panel:} classifications for H-ATLAS detected
    ETGs from A13/A15. \textbf{Bottom panel:} classifications for H-ATLAS
  non-detected galaxies as a control.}\label{figure:mrphhists}
\end{figure}

The dusty ETGs from A13/A15 studied here have been visually classified by both
the GAMA team \citep{kelvin14} and the SAMI team using essentially the
same classification criteria. Two of the key differences between the
classifications are that they are made up of independent groups of
classifiers and they used different images in the classification process. GAMA
classifications of \citet{kelvin14} are based on false colour $g$,
$i$, and $H$ band composite images while SAMI classifications employ
SDSS DR9 $gri$ images. The use of longer wavelength data has
resulted in the GAMA classifications tending somewhat towards
earlier types. There is likely also an influence of the third key difference between SAMI
and GAMA classifications, namely those of SAMI include signs of star
formation (based on galaxy colour rather than morphology alone) to
distinguish LTGs from ETGs. This would help to explain why such a
large number of A13/A15 ETGs are identified as Sa galaxies, which are
morphologically difficult to separate from S0 galaxies beyond $z$ =
0.05, but would be identified by SAMI as later types due to their blue colours.
This difference is illustrated in the top panel of Figure \ref{figure:mrphhists}
where we show the GAMA and SAMI classifications for the A13/A15 dusty ETGs
studied here. We show classification histograms for H-ATLAS
non-detections in the bottom panel of Figure \ref{figure:mrphhists}
and, although GAMA classifications are still slightly skewed towards
earlier types, the level of agreement is improved compared to H-ATLAS
detections. 

There is often an inherent
assumption that there is a connection
between visual classification of a galaxy as an ETG and the presence
of a hot, X-ray emitting ISM
\citep[e.g.][A13/A15]{rowlands12,martini13}. This may not be fully justified, and, in the case of A13/A15,
the inclusion of a large number of Sa galaxies makes this connection
more dubious. Comparing those GAMA ETGs containing dust to those that
do not, A13 show that dusty ETGs
are bluer, less concentrated, and have lower S\'{e}rsic
indices. Further, A13 dusty ETGs have NUV-r colours more similar to
H-ATLAS detected GAMA LTGs than to non-detected ETGs. Thus, from A13
there is already an indication that many H-ATLAS detected visual ETGs
from GAMA have properties more like LTGs than giant ellipticals hosting X-ray halos. 

The kinematic analysis of the 49 A13/A15 H-ATLAS detections with reliable SAMI
observations agrees well with this assessment. In Section
\ref{section:detgkine} we show that 44/49 of these galaxies have a
$V_{c}$ versus $\sigma_{mean}$ in Figure \ref{figure:agiuskcl}
suggesting kinematics largely dominated by rotation (termed RDGs here). All of these
galaxies have $\sigma_{mean}$ below 150 km s$^{-1}$ (with only 6 above
$\sigma_{mean}$ = 100 km s$^{-1}$), the approximate
value above which galaxies have X-ray luminosities exceeding the value
expected for the cumulative emission from supernova remnants and X-ray
binaries in empirical studies
\citep{boroson11,sarzi13,kim15,goulding16}.

It has also been shown by \citet{anderson15} that $M_{*}$ can often be
used to identify galaxies hosting an X-ray emitting ISM, with the
clearest evidence found for galaxies with log$_{10}$($M_{*}$) > 10.8
$M_{\odot}$. In fact, it is likely that $M_{*}$ is more fundamental in
determining the presence of such a hot ISM as it is the high mass
concentration of these galaxies that prevents hot gas from escaping
into the inter-galactic medium. This means that $\sigma_{mean}$ is a
secondary indicator arising through the relationship between $M_{*}$ and $\sigma$ in
ETGs \citep{faber76}. This connection occurs because massive ETGs derive their
dynamical support from random motions, which is not the case for
galaxies with significant rotation. 

Assuming $M_{*}$ is a better
indicator for the presence of hot X-ray emitting gas than
$\sigma_{mean}$ in rotating galaxies, we also identify those H-ATLAS
detected RDGs with log$_{10}$($M_{*}$) > 10.8 in Figure
\ref{figure:agiuskcl}. 41/44 H-ATLAS detected RDGs are found to have
masses below this limit, supporting our assertion that they do not
host a hot ISM. Although the remaining three RDGs are massive enough
to host an X-ray emitting halo, they are also found to
have rapid rotation, with $V_{c}$ > 160 km
s$^{-1}$ in all three cases. \citet{negri14a} and \citet{negri14b} have shown that rapid
rotation allows massive galaxies to host a disk of cold gas even
in the presence of an X-ray emitting halo. This means that, regardless
of the properties of the ISM in these three galaxies, dust residing in
their disks can be long lived, thus an external origin for their dust
content is unnecessary. 

Thus, the visual classification of a galaxy as an ETG should not be
assumed as clear evidence for the presence of a hot ISM that is
inhospitable to dust, in agreement with previous works
\citep[e.g.][]{sansom00,sansom06}. On the contrary, despite their
appearance many visually-classified ETGs are actually
rotationally-supported, disk-like, star-forming galaxies with a relatively normal
dust content comparable to galaxies found on the star-forming main
sequence. In other words, the term early-type in this case does not
imply structural difference, but mainly a difference in colour and possibly
SFR. This means that the dust content of these galaxies is likely
produced internally through normal processes such as supernovae and
stellar winds without the need for external mechanisms such as
cooling flows or minor mergers. 

\subsection{Dispersion Dominated Galaxies, Dust, and Merger Rates}\label{section:dustdisc}

From our sample of 49 dusty ETGs from A13/A15 with reliable SAMI
observations, we identify 4 galaxies in Figure \ref{figure:agiuskcl}
that are kinematically consistent with
being dispersion-dominated systems. SDSS DR9 images of these
galaxies are shown in Figure \ref{figure:agellip} alongside velocity
and velocity dispersion maps from the SAMI galaxy survey. All four galaxies
exhibit inconsistencies between their stellar and ionised gas
indicating recent stochastic processes such as gas accretion through merging. Galaxies 551505 and 534655 are
found to have kinematic misalignments of $\sim$90$^{\circ}$ while
508180 and 511892 have stellar to gas $V_{c}$ ratios < 0.60,
inconsistent with the range observed for asymmetric drift in LTGs
\citep[0.75-0.89][]{martin13,cortese16}. Furthermore, the regular
appearance of these galaxies from SDSS observations suggests that if
mergers are responsible for these kinematic inconsistencies then these
mergers must either be minor,
as major mergers typically result in disturbed morphologies
\citep[e.g.][]{larson16}, or they occurred in the fairly distant past thus
having allowed significant time for dynamical relaxation. A possible caveat, however, is that
observations deeper than those from SDSS may reveal disturbed
morphologies apparent as low surface brightness features
\citep[e.g.][]{sheen12}. 

Evidence that the dust content of these four galaxies may have
been recently accreted comes by comparing $M_{d}$ to other
galaxy properties. First, Figure \ref{figure:mdsfr} shows that DDGs
have suppressed SFR compared to normal star-forming
galaxies, a feature seen in simulations of wet minor mergers
\citep{peirani10,davis15,gereb16}. DDGs are also
significantly offset above the $M_{d}$-$M_{*}$ relationship for
star-forming galaxies shown in Figure \ref{figure:mdms}, similar to
dusty ETGs from \citet{dariush16} who observe little variation in
$M_{*}$ with varying $M_{d}$. This lack of a correlation is suggested
as further evidence of external accretion. Indeed, extremely dust poor
massive elliptical galaxies will fall far above the
$M_{d}$-$M_{*}$ trend for star-forming galaxies, occupying the top
left region of Figure
\ref{figure:mdms}. A subsequent wet minor merger will provide a
negligible increase in $M_{*}$ while significantly increasing $M_{d}$,
thus the merger remnants will move horizontally to the right in Figure
\ref{figure:mdms} towards the region occupied by kinematic ETGs
discussed here. 

Assuming all four H-ATLAS detected DDGs in this work have acquired
their dust content in a merger, how do our results compare with
expectations based on the cosmological rates of mergers at low
redshift? \citet{martini13} compare the measured rate of minor
mergers to the theoretical
estimates of the destruction time for dust in hot gas of
\citet{draine79}. Following \citet{stewart09}, the
authors make a rough prediction of the expected fraction of dusty
ETGs, $f_{dust}$, based on estimates of the merger rate of $R_{merg} =
0.07-0.2$ Gyr$^{-1}$ and a dust lifetime of $\tau_{dust}$ < 0.02 Gyr
following
\begin{equation}
  f_{dust} = R_{merg}\tau_{dust}
\end{equation}
This gives $f_{dust}$ < 0.14-0.4\%
implying that a purely external accretion
scenario for large samples of dusty ETGs is extremely unlikely from a
statistical perspective. The results of our study provide a possible
solution to this tension. First we note that the fraction of ETGs with
dust quoted by \citet{martini13} is 0.6, whereas for GAMA ETGs in A13
there is only a 29\% detection rate from the H-ATLAS survey. As we
have shown, however, many of these galaxies are kinematically
inconsistent with the presence of a hot ISM. Among DDGs, we find an
even lower H-ATLAS detection rate of 11\% (4/35), a factor of
$\sim$5 lower than the value assumed by \citet{martini13}, yet still
significantly larger than their predicted value of $f_{dust}$ <
0.14\%. 

The argument of \citep{martini13}, however, is dependent on a number
of assumptions regarding the timescales and conditions of dust
accretion in ETGs. Some works
have estimated a timescale for gas stripping on the order of a few
times 10$^{8}$ yr \citep{takeda84,murakami99}, which could further
increase $f_{dust}$ predictions by an order of magnitude to
$\sim$1.4-4.0\%. The remaining tension between this estimate and
the estimate of $f_{dust}$=11\% found in this work may be partially due to
incompleteness as a result of our small sample size. Another possibility,
though, is the recent suggestion that
accretion of dust that is embedded in a larger cold medium may be
shielded from the harsh ISM resulting in a further significant
increase in dust lifetimes
\citep[e.g.][]{clemens10,dasyra12,finkelman12}. A full understanding
of just how much longer dust may survive in such a scenario
is beyond the scope of this work. It should also be noted that, in
those galaxies with $V_{c,star}$/$V_{c,gas}$ < 0.6, the fact that the
gas is kinematically aligned with the stars suggests that if the
presence of the dust is truly the result of a merger then it must have
had sufficient time to undergo dynamic relaxation. This process should
occur on Gyr timescales \citep{lake83,davis16}, supporting the idea
that the direct, sub-galactic, environment of dust in ETGs may
drastically increase the lifetime of interstellar dust. 

We can also discuss the comparison between stellar and ionised gas
kinematics of dust-free DDGs in this study, i.e. H-ATLAS
non-detections from A13/A15. Of the 31 H-ATLAS non-detected DDGs
included here, 36\% (11/31) show signs of kinematic discrepancies
between their ionised gas and stars. Among these 11, 7/11 are
kinematically misaligned while 4/11 are aligned with significantly larger gas
$V_{c}$ compared to that of the stars. The remaining 20 galaxies,
however, do not have secure enough detections of ionised gas to
provide clean ionised gas kinematics maps. Considering our entire sample of 540 SAMI galaxies
with high quality stellar kinematics observations, we find a total subsample of
71 DDGs. Out of these, 49\% (35/71) show kinematic irregularities,
however only 58\% (41/71) have strong ionised gas detections. Among
the 35 kinematically irregular DDGs, 22 are kinematically misaligned
while the remaining 13 have aligned ionised gas rotating significantly faster
than the stars. Our finding that 54\% (22/41) of DDGs with appreciable
amounts of ionised gas are kinematically misaligned agrees well with
the findings of Bryant et al. (in preparation) who find that, among a
larger sample of SAMI galaxies, 40-53\% of ETGs with high
signal-to-noise ionised gas emission exhibit kinematically misaligned
gas. 

This begs the question, how are kinematically irregular, H-ATLAS
non-detected DDGs related to DDGs with H-ATLAS detections? The
strongest statement we can make in this regard is that the presence of
dust does not have a large impact on the relative dynamics of gas and
stars in DDGs with significant gas. As we do not have strong
constraints on the possible dust content of H-ATLAS non-detected
galaxies, unlike H-ATLAS detections, we do not have the secondary
indications of a mergers as an explanation for their kinematic
irregularity (e.g. $M_{d}$ vs $M_{*}$). We can simply say that the
presence of ionised gas is likely associated with a stochastic process
that has also affected the gas kinematics in these galaxies.
Unlike dust, however, ionised gas is not always directly associated with cold
gas in galaxies. Indeed, a number of works have shown that old stellar
populations (such as post-AGB stars), active galactic nuclei (AGN), or
even interactions between warm and hot (X-ray emitting) gas phases
may be the dominant sources of ionised gas emission in ETGs
\citep[e.g.][]{binette94,sarzi10,yan12}. This means that the detection
of ionised gas emission in the absence of a secure detection of cold
dust is not inconsistent with the presence of a hot, X-ray emitting
ISM. 

A caveat here, however, is that some H-ATLAS non-detected
galaxies from A13/A15 (particularly those DDGs with strong ionised gas
emission) may contain appreciable amounts of dust, but have far-IR
fluxes below the sensitivity limits of H-ATLAS. Assuming the presence
of ionised gas is \textit{always} associated with dust in our sample of SAMI galaxies
would increase our estimate of the number of DDGs with dust to 58\%
(41/71). Thus the level of tension between merger rates and dust
lifetimes in ETGs here would roughly match that seen by
\citet{martini13}. This assumption, however, is unfounded and, as we
have noted, estimates of dust lifetimes in the presence of a complex,
multiphase ISM in ETGs are equally uncertain. Further study of the ISM
of massive galaxies and the precise conditions of gas accretion onto
these systems will be required to understand the underlying cause of
this tension.


Finally, we note that out of the 41 DDGs from our full SAMI sample
that exhibit appreciable ionised gas emission, 5 do not show clear
discrepancies between ionised gas and stellar kinematics. These
galaxies, however, have not been observed by H-ATLAS, thus the
presence of dust is uncertain. Given the above discussion regarding
sources of ionising radiation in ETGs, unless these galaxies can be
shown to host cold gas or dust, they should not be considered to be
peculiar objects.

\section{Summary}\label{section:conclusions}

In this paper we have analysed the 2D kinematics of visually
classified ETGs from \citet{agius13} and \citet{agius15} (A13/A15) using IFS data from the SAMI Galaxy
Survey. The sample of A13/A15 includes 220 H-ATLAS detected (dusty)
galaxies and 551 H-ATLAS non-detected (dust-free) galaxies. We begin
by measuring the stellar circular velocity, $V_{c}$, and
flux-weighted, global, stellar velocity dispersion, $\sigma_{mean}$, for a sample of 540 SAMI
galaxies for which we can measure $V_{c}$ beyond the turnover radius of
the rotation curve. These values provide a kinematic selection designed to determine those visual
ETGs that are consistent with having dispersion-dominated stellar
kinematics, indicative of the presence of a hot, X-ray emitting ISM. This
selection is then applied to visually classified ETGs from A13/A15
that have currently been observed with SAMI. Finally, we examine the
dust properties of these galaxies in comparison with our kinematic
selection in order to better understand the origin of the dust in
these systems. Our key results are as follows:
\begin{itemize}
  \item Selecting A13/A15 ETGs based on $V_{c}$ and
    $\sigma_{mean}$ we find 11\% (4/35) of dispersion-dominated
    A13/A15 galaxies are H-ATLAS detected. This is in contrast to the 29\%
    (220/771) detection rate for the full A13/A15 sample and 40\%
    (45/113) for rotation-dominated galaxies. Thus the detection of
    dust in visually classified ETGs is 3.5$\times$ more likely in
    galaxies with disk-like rotation than in those with kinematics
    more consistent with massive elliptical galaxies.
  \item Similarly, only 8\% (4/49) of H-ATLAS detected ETGs from A13/A15 with SAMI
    observations are kinematically consistent with being true,
    dispersion-dominated galaxies. The remainder have kinematics more
    similar to blue, visually-classified LTGs and S0/Sa galaxies.
  \item 100\% of these dispersion-dominated, dusty ETGs
    exhibit inconsistencies between their stellar and ionised gas
    kinematics suggestive of recent merger activity and an external
    origin for their dust content. The corresponding rate of gas
    versus stellar kinematic discrepancies in our full sample of
    dispersion-dominated SAMI galaxies is 45\% (34/75).
  \item The four dispersion-dominated, dusty ETGs in our sample are
    also extremely massive and thus quite likely to host a hot, X-ray
    emitting halo. As such, external accretion scenario is the most
    viable source for their dust content. Observations of a suppressed
    star formation in these four galaxies, typical of gas accreted
    onto massive galaxies \citep[e.g.][]{gereb16}, further supports this
    assertion.
  \item The low velocity dispersions as well as low masses and/or
      rapid rotation of the remaining galaxies, suggest that dust in these
    systems may be long-lived thereby eliminating any need for an external 
    scenario for the origin of their dust content. 
\end{itemize}
We have shown that these results may help to reduce the tension between expected
dust lifetimes in massive ETGs \citep{draine79,clemens10}, observed merger rates
\citep{lotz11}, and the observed number of visual ETGs containing dust
\citep{martini13}. A more complete understanding of the complex,
multiphase ISM of massive ETGs, as well as the exact conditions
through which gas is accreted onto these systems, will be necessary in
understanding this tension fully. 

\vspace{.75cm}
RB acknowledges support under the Australian Research Council's (ARC) Discovery Projects funding scheme (DP130100664).
JvdS is funded under Bland-Hawthorn's ARC Laureate Fellowship
(FL140100278). SMC acknowledges the support of an Australian Research
Council Future Fellowship (FT100100457). SB acknowledges the funding
support from the Australian Research Council through a Future
Fellowship (FT140101166). Support for AMM is provided by NASA through
Hubble Fellowship grant \#HST-HF2-51377 awarded by the Space Telescope
Science Institute, which is operated by the Association of
Universities for Research in Astronomy, Inc., for NASA, under contract
NAS5-26555. M.S.O. acknowledges the funding support from the
Australian Research Council through a Future Fellowship Fellowship
(FT140100255). We would also like to thank the anonymous referee for
comments and suggestions that have improved the clarity and
readability of this work. The SAMI Galaxy Survey is based on observations made at
the Anglo-Australian Telescope. The Sydney-AAO Multi-object Integral
field spectrograph (SAMI) was developed jointly by the University of
Sydney and the Australian Astronomical Observatory. The SAMI input
catalogue is based on data taken from the Sloan Digital Sky Survey,
the GAMA Survey and the VST ATLAS Survey. The SAMI Galaxy Survey is
funded by the Australian Research Council Centre of Excellence for
All-sky Astrophysics (CAASTRO), through project number CE110001020,
and other participating institutions. The SAMI Galaxy Survey website
is http://sami-survey.org/. GAMA is a joint European-Australasian project based
around a spectroscopic campaign using the Anglo-Australian
Telescope. The GAMA input catalogue is based on data taken from the
Sloan Digital Sky Survey and the UKIRT Infrared Deep Sky
Survey. Complementary imaging of the GAMA regions is being obtained by
a number of independent survey programmes including GALEX MIS, VST
KiDS, VISTA VIKING, WISE, Herschel-ATLAS, GMRT, and ASKAP providing UV
to radio coverage. GAMA is funded by the STFC (UK), the ARC
(Australia), the AAO, and the participating institutions. The GAMA
website is http://www.gama-survey.org/.




\bibliographystyle{mnras}
\bibliography{refs} 



\appendix

\section{Dust Mass Upper Limits for H-ATLAS
  Non-Detections}\label{appendix:dustul}

As described in Section \ref{section:dustmass}, $M_{d}$ for H-ATLAS
detected galaxies is measured by the GAMA survey team using the
\texttt{MAGPHYS} spectral fitting code of
\citet{dacunha08}. \texttt{MAGPHYS} relies on detections in the far-IR in
order to provide reliable estimates of $M_{d}$, therefor those
$M_{d}$ values provided by \texttt{MAGPHYS} for H-ATLAS non-detected galaxies are
highly uncertain. Thus, for these galaxies we estimate upper limits on
the dust masses in the following manner:

First we take
the upper limit for the flux in the 250 $\mu$m \textit{Herschel} SPIRE
band, $F_{250}$, to be $\sim$33 mJy \citep{dunne11}. This is then converted to an
upper limit on $M_{d}$ using \citep{hildebrand83}:
\begin{equation}\label{equation:dustmass}
  M_{d}=\frac{F_{250}D_{L}^{2}K_{250}}{\kappa_{250}B(1+z)(T_{d})_{250}}
\end{equation}
where $D_{L}$ is the luminosity distance to each galaxy, computed from
spectroscopic redshifts, $\kappa_{250}$ is the mass absoprtion
coefficient assumed to be 0.89 m$^{2}$ kg$^{-1}$ at 250 $\mu$m
  \citep{dunne11}, and $B(T_{d})_{250}$ is the Planck function at 250
  $\mu$m and at  dust temperature $T_{d}$. For all non-detected galaxies we
  simply fix $T_{d}$ at 22.1 K, the average value computed through
  grey-body fitting for H-ATLAS detected sources from
  A13. Equation \ref{equation:dustmass} also includes a factor of
  (1+$z$) and a $K$-correction, which is given by:
\begin{equation}
  K =\left( \frac{\nu_{obs}}{\nu_{rf}}\right)^{3+\beta} \frac{e^{(h\nu_{rf}/kT_{d})}-1}{e^{h\nu_{obs}/kT_{d}}-1}
\end{equation}
where $\nu_{obs}$ and $\nu_{rf}$ are the observed and rest-frame
frequency, $\beta$ is the dust emissivity index \citep[fixed at
1.5][]{dye10}, $h$ is the Planck constant, and $k$ is the Boltzmann
constant. $M_{d}$ upper limits are depicted along side
\texttt{MAGPHYS} dust masses for H-ATLAS
detections in Section \ref{section:dustprops}.

\section{Rotation Curve Extraction and Measurement of
  $V_{C}$}\label{appendix:vc}

\begin{figure}
  \includegraphics[width=\columnwidth]{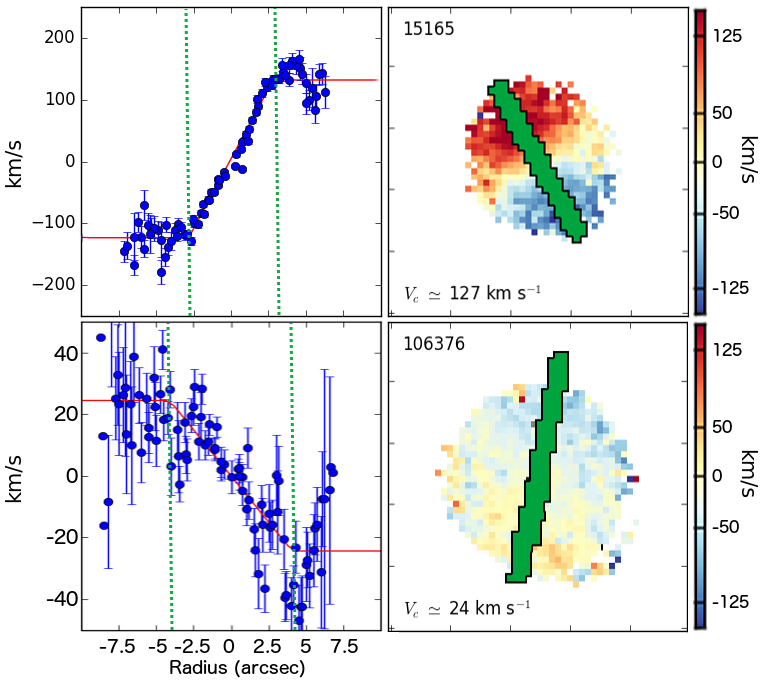}
  \caption{Two examples of our fitting procedure for $V_{c}$. The
    right column illustrates, in green, the artificial slits that are traced
    along the kinematic major axis of our stellar velocity maps (box
    size = 25$\farcs$0) and
    the left column shows as blue circles the individual velocities of
  pixels within the slit. Also shown in the left column as solid red
  lines are rotation curve models for each galaxy given by Equation
  \ref{equation:rcmod}, and the green dotted lines indicate the
  locations of the turn-over radii, $r_{t}$. We have chosen these two
  examples to show a
  rapidly rotating galaxy with a well defined rotation curve (top row)
  as well as a much more slowly rotating galaxy (bottom row) roughly
  at the two extremes of our $V_{c}$ measurements.}\label{figure:vcf_ex}
\end{figure}

After measuring the kinematic PA of our kinematics maps, we use this
value as input in extracting the rotation curves for each galaxy.
This is done by first determining the $x$ and $y$ positions of the
galaxy centre from the stellar flux maps, $F(x,y)$ using:
\begin{equation}\label{equation:rcmod}
  x_{c}= \frac{\sum_{i} \sum_{j} i \times F(i,j)}{\sum_{i} \sum_{j} F(i,j)}\quad ;\quad y_{c}=\frac{\sum_{i} \sum_{j}j \times F(i,j)}{\sum_{i} \sum_{j}F(i,j)}
\end{equation}
where $x_{c}$ and $y_{c}$ are the x and y positions of the galaxy
centre. An artificial slit with a width of 1$\farcs$5 (3 spaxels) is traced
across the stellar velocity map with its position defined by the
measured PA and galaxy centre. Examples of these artificial slits are
shown in green in the right column of Figure \ref{figure:vcf_ex}. The radius, $r(x,y)$, and velocity,
$v(x,y)$, are recorded at each spaxel within the slit. Here we define
$r(x,y)$ $=$ $\sqrt{(x-x_{c})^{2}+(y-y_{c})^{2}}$ with the sign taken
to match the sign of $x-x_{c}$. The choice of the definition of
positive and negative radii is arbitrary, however, as this simply defines a
positive or negative $V_{c}$. In the end, the final value of the
circular velocity is taken as $|V_{c}|$. 

The stellar rotation curves extracted in this way are then used to determine
the rotation velocity following the procedure of
\citet{epinat09}. This is done by fitting a piecewise function of the
form:
\begin{equation}\label{equation:vrfit}
   V(r) = \left\{
     \begin{array}{lr}
      -V_{c}, & r \leq -r_{t}\\
       V_{c}(r/r_{t}), & -r_{t} < r < r_{t}\\
       V_{c} & r_{t} \leq r
     \end{array}
   \right.
\end{equation}
where $r_{t}$ is the turn-over radius of the rotation curve that
defines where the flat portion of the rotation curve begins. This
value is left as a free parameter. Examples of these fits for galaxies
with low and high (apparent) $V_{c}$ are shown in the left column of
Figure \ref{figure:vcf_ex}. Although a fairly common feature of
galaxy rotation curves is a decline in $V_{c}$ beyond the turnover
radius, $r_{t}$ \citep[e.g.][]{deblok08}, the coverage of our datacubes often does not
extend to large enough radii to capture this behaviour. Thus we choose
to use the relatively simple model given in
Equation \ref{equation:vrfit} as including more parameters is more
likely to result in spurious fits, particularly for low S/N spaxels at
large radii. 

This procedure has the inherent
assumption that the $\sim$8$\farcs$0 covered by SAMI datacubes is
larger than $r_{t}$, which is not true for many galaxies. We check whether or not observations of each galaxy
extend beyond $r_{t}$ to test this. For each
galaxy we first identify the spaxel in our traced slit that is
furthest from $(x_{c},y_{c})$. If the radius measured for this spaxel
is larger than $r_{t}$ then we flag this as a reliable measurement
of $V_{c}$. We also visually inspect the rotation curve fits for each
galaxy flagged as reliable for verification, and in this
process, a small number of galaxies were identified with spurious fits
and subsequently flagged and removed from our sample. Finally, some SAMI observations include a low number of
spaxels with high signal-to-noise data, which will reduce the
reliability of our fitted value of $V_{c}$. We perform a test using
galaxies with high fidelity rotation curves in
which we incrimentally reduce the artificial slit length by one spaxel
and remeasure $V_{c}$. We find that for slits containing more than 30
spaxels we are able to recover $V_{c}$ measured from the full slit for
>95\% of galaxies tested. This fraction falls off rapidly below 30
spaxels, thus galaxies for which slits contain less than 30 spaxels
are excluded from our analysis. Among
the 753 galaxies tested, we find that 554 galaxies meet these 
requirements.

We next 
correct our stellar $V_{c}$ measurements for the effects of inclination, which causes observed
$V_{c}$'s to be lower than the intrinsic rotation velocity of a given
galaxy. First we determine the inclination of each galaxy using
\begin{equation}
  \textrm{cos}^{2}i = \frac{(1-\epsilon)^{2}-\alpha^{2}}{1-\alpha^{2}}
\end{equation}
where $i$ is the galaxy inclination, $\epsilon$ is the observed
ellipticity, and $\alpha$ is the intrinsic flattening for a given
galaxy. For each galaxy we provide a rough estimate of
$\alpha$ based on the following criteria: first we separate galaxies
into those that are strongly disk-dominated from those with a
significant influence from a central bulge. The former are identified
as having bulge to total ratios (B/T) less than 0.3 while the latter
have B/T larger than 0.3, which roughly follows the findings of
\citet{graham08}. Here we take B/T from $r$-band values of \citet{simard11} who perform
2D bulge+disk decompositions for SDSS galaxies using the GIM2D software
\citep[v3.2][]{simard02}. For disk-dominated galaxies we fix $\alpha$
at 0.23, which is the average value found when comparing $\alpha$ values
reported for spiral galaxies by \citet{lambas92} and
\citet{padilla08}. Galaxies with B/T > 0.3 are then separated into
pure ellipticals and S0/Sa galaxies based on SAMI morphological
classifications. We assign an $\alpha$ of 0.55 to the S0 and Sa classes
\citep{lambas92,noordermeer06} and a value of 0.63 to elliptical
galaxies, the former being the average value found for slow rotators
in the ATLAS$^{\textrm{3D}}$ survey \citep{weijmans14}. This large
$\alpha$ value for elliptical galaxies is appropriate because these
object appear relatively round even when the viewing angle is
perpendicular to the axis of rotation. Galaxies with low intrinsic
rotation and a spheroidal shape, for example, would have $V_{c}$
significantly over estimated if it is assumed the galaxy is much
flatter. Thus adopting $\alpha$=0.63 for elliptical galaxies provides
a conservative $V_{c}$ correction appropriate for dispersion-supported
galaxies. The exact assumptions regarding $\alpha$ will have only a
minor effect on our results as this value is used to correct $V_{c}$
while, as we will show in Section \ref{section:kineclass}, our
kinematic selection is primarily based on stellar velocity dispersion.

Finally, we compute
the inclination corrected stellar $V_{c}$, $V_{c,corr}$, as
\begin{equation}
  V_{c,corr}=\frac{V_{c}}{(1+z)\textrm{sin}i}
\end{equation}
This correction inherently assumes that galaxies observed face-on are
perfectly circular, which is certainly not accurate for all
galaxies. By construction, this process has a relatively small effect
on ETGs while LTGs may have $V_{c}$ underestimated by up to 270 km
s$^{-1}$. This is typically the case galaxies observed close to face-on, for which
$V_{c}$ is already quite uncertain, however. Among the 563 galaxies with well
sampled rotation curves, we find a median increase in $V_{c}$ due to
our inclination correction of 4.7 km s$^{-1}$, and only 7\% have an
increase in $V_{c}$ of more than 50 km s$^{-1}$. As we are interested
in ETGs in this work, cases such as this will not affect our results. 

\section{$\sigma_{mean}$: Masking and Beam Smearing
  Correction}\label{appendix:bs}

In this appendix we describe in detail our methods of masking bad
spaxels prior to measuring $\sigma_{mean}$ and correcting
$\sigma_{mean}$ for the effects of seeing, commonly referred to as
beam smearing.

We define
bad spaxels as those that do not satisfy $\sigma_{error} < \sigma
\times 0.1 + 25$ km s$^{-1}$. This requires the measured error in $\sigma$ in
each spaxel to be smaller than a fraction of the measured
$\sigma$. The inclusion of the $+25$ km s$^{-1}$ is needed in order
that we do not exclude a majority of spaxels with a low measurement of
$\sigma$. Finally, we also exclude spaxels with $\sigma$ < 35 km
s$^{-1}$, which is the limit to which we trust our measurements
\citep[see][for more details]{vandesande17}, for more on tests of our pPXF procedure). 
$\sigma_{mean}$ is then measured as the flux weighted velocity dispersion
over unmasked spaxels in our stellar velocity dispersion maps. Formally this is
defined as:

Our measurements of $\sigma_{mean}$ employ \textbf{all} spaxels meeting the
quality cut described above. We test the robustness of $\sigma_{mean}$
by remeasuring this value within radii between 1$\farcs$0 and
8$\farcs$0 (2-16 spaxels). We find that $\sigma_{mean}$ measurements
level off beyond 1$\farcs$5, and remain unchanged out to
8$\farcs$0. This means that $\sigma_{mean}$ measurements are robust
for all galaxies that meet our $V_{c}$ quality cut, i.e. velocities
are measured beyond $r_{t}$. 

The major difficulty in estimating the global velocity dispersions
from IFS observations is accounting for the effects of beam smearing,
which can artificially inflate $\sigma$ measured in
individual spaxels \citep[e.g.][]{pracy05,law09}. This effect is enhanced in the central regions of
rapidly rotating galaxies where large velocity gradients are observed
over individual spaxels. Beam smearing is a complex effect, acting in
all three dimensions of IFS datacubes, and a significant ongoing
effort to understand beam smearing in SAMI data is underway. In the
meantime, we perform a simple beam smearing correction on $\sigma_{mean}$
following \citet{bassett14}. For each galaxy we estimate the
additional $\sigma$ induced by beam smearing, $\sigma_{bs}$, as:
\begin{equation}
  \sigma_{bs} \approx \frac{dV}{d\theta}\sigma_{\theta}
\end{equation}
where $dV$ is the velocity gradient defined by our $V_{c}$ fits as $V_{c}/r_{t}$ (see
\ref{section:vc}), $d\theta$ is the spaxel size of 0$\farcs$5, and $\sigma_{\theta}$
is the seeing of our observations. Seeing values for SAMI galaxies are
catalogued at the time of observation with the median seeing at the
AAT site being 1$\farcs$8. We estimate a ``beam
smearing corrected'' $\sigma_{mean}$ as $\sigma_{m,corr} =
\sqrt{\sigma_{mean}^{2}-\sigma_{bs}^{2}}$. The typical corrections
measured in this way reduce the measured $\sigma_{mean}$ by $\sim$0-30 km
s$^{-1}$ with a clear dependence on $V_{c}$. Below $V_{c}$ = 60 km
s$^{-1}$, corrections are closer to $\sim$0-5 km s$^{-1}$. Note that
this may result in measured values of $\sigma_{m,corr}$ below 35 km
s$^{-1}$, which we regard as the lower limit to which we trust
measurements of the stellar $\sigma$ in individual spaxels. Prior to
performing our beam smearing correction, all measured $\sigma_{mean}$
values are larger than 40 km s$^{-1}$, thus $\sigma_{m,corr}$ values
below 35 km s$^{-1}$ are entirely resultant from our beam smearing
correction. 

We note that rotation curves may also be affected by beam smearing,
in particular the velocity gradient in the central regions. Our
$V_{c}$ measure is largely constrained by the asymptotic velocities at
large radii and thus will be minimally affected by beam smearing (if at
all). We deem the $\sigma_{mean}$ correction described here necessary,
however, as this is a flux weighted quantity, is biased
towards the central regions where beam smearing effects are at a maximum.


\bsp	
\label{lastpage}
\end{document}